# Microshift: An Efficient Image Compression Algorithm for Hardware

Bo Zhang, *Student Member, IEEE,* Pedro V. Sander, Chi-Ying Tsui, *Senior Member, IEEE,* and Amine Bermak, *Fellow, IEEE*

***Abstract*—In this paper, we propose a lossy image compression algorithm called Microshift. We employ an algorithm-hardware co-design methodology, yielding a hardware friendly compression approach with low power consumption. In our method, the image is first micro-shifted, then the sub-quantized values are further compressed. Two methods, FAST and MRF model, are proposed to recover the bitdepth by exploiting the spatial correlation of natural images. Both methods can decompress images progressively. On average, our compression algorithm can compress images to 1.25 bits per pixel with a resulting quality that outperforms state-of-the-art on-chip compression algorithms in both peak signal-to-noise ratio (PSNR) and structual similarity (SSIM). Then, we propose a hardware architecture and implement the algorithm on an FPGA. The results on the ASIC design further validate the low hardware complexity and high power efficiency, showing our method is promising, particularly for low-power wireless vision sensor networks.**

## I. Introduction

Traditional image compression algorithms are designed to maximize the compression rate without significantly sacrificing the perceptual quality [1], [2]. For wireless vision sensor networks such as smart home surveillance and smart pills, power consumption is also an important design factor [3]. However, many complex compression standards, though highly efficient in terms of compression performance, are unsuitable for such application scenarios [4], [5]. In recent years, some on-chip algorithms have been proposed [6]. However, while these works target low hardware complexity, their compression performance is usually compromised.

In this work, we are interested in the compression algorithm for WVSNs, where the image data is acquired at the sensor node, and then wirelessly transmitted to the central node for processing. Unlike general image compression algorithms, designing a compression algorithm in WVSNs should consider the tradeoff between compression ratio, image quality and also implementation complexity [7]–[9]. Specifically, the compression algorithm should have the following features:

*a) Hardware friendliness:* The operators should be easy to implement. For example, floating-point calculations should not be used, single raster scan of images is preferred, and the memory usage should be minimized. To meet all of these requirements, the algorithm has to be designed with hardware implementation considerations.

*b) High compression ratio:* The compression ratio should be high enough so that the transmission throughput is reduced, thus saving power.

*c) Unbalanced complexity:* The central node which receives the data can have strong computational capability. Therefore, while we must minimize the compression complexity at the sensor node, the decompression complexity is not a major concern.

*d) Progressive decompression:* Since wireless transmission is power hungry, the decompression is expected to be progressive. That is, if the information resulted from the coarse decompression is found unimportant, the central node can notify the sensor to early terminate the transmission for that frame for power savings.

*e) High image quality:* Finally, all of the above must be achieved without significantly compromising the visual quality of the resulting image.

Under these guidelines, we propose *Microshift*, which achieves good compression performance and can be easily integrated with the modern CMOS image sensor architecture. The compression has two major steps. Inspired by [10], the pixel values are initially shifted by a $3 \times 3$ microshift pattern, and then these values are sub-quantized with fewer bits. By taking advantage of the spatial correlation between adjacent pixels, the original bitdepth can be effectively recovered. This first step is lossy. In the second step, subimages containing pixels that share the same microshift are losslessly compressed through either intra- or inter-prediction. The overview of the entire framework is illustrated in Fig. 1.

The decompression is performed in a reverse manner. We propose two methods for bitdepth reconstruction from the micro-shifted image. The first approach infers the value for each pixel according to its neighboring pixels. We call this method FAST since it runs efficiently. The weighted least square (WLS) filter is used to further suppress artifacts [11]. The second approach is based on Markov random field (MRF) optimization, which estimates the pixel values based on their maximum a posteriori (MAP) probability. Through global optimization, the MRF decompression model provides better image quality at the cost of slower computational speed. Both FAST and MRF methods can decompress the image progressively.

We tested our method on standard images and show that our algorithm outperforms other on-chip compression methods. On the testing dataset, the average bit per pixel (BPP) after compression is 1.25, the peak signal-to-noise ratio (PSNR) is 33.16 dB and the structural similarity (SSIM) index is 0.902. To validate the low hardware complexity of our method, we propose a hardware architecture achieving power FOM (power normalized by the frame rate and the number of pixels) as low as 19.7 pW/pixel·frame. When comparing with the state-of-the-art algorithms, our method shows the best trade-off between power consumption and compression performance.

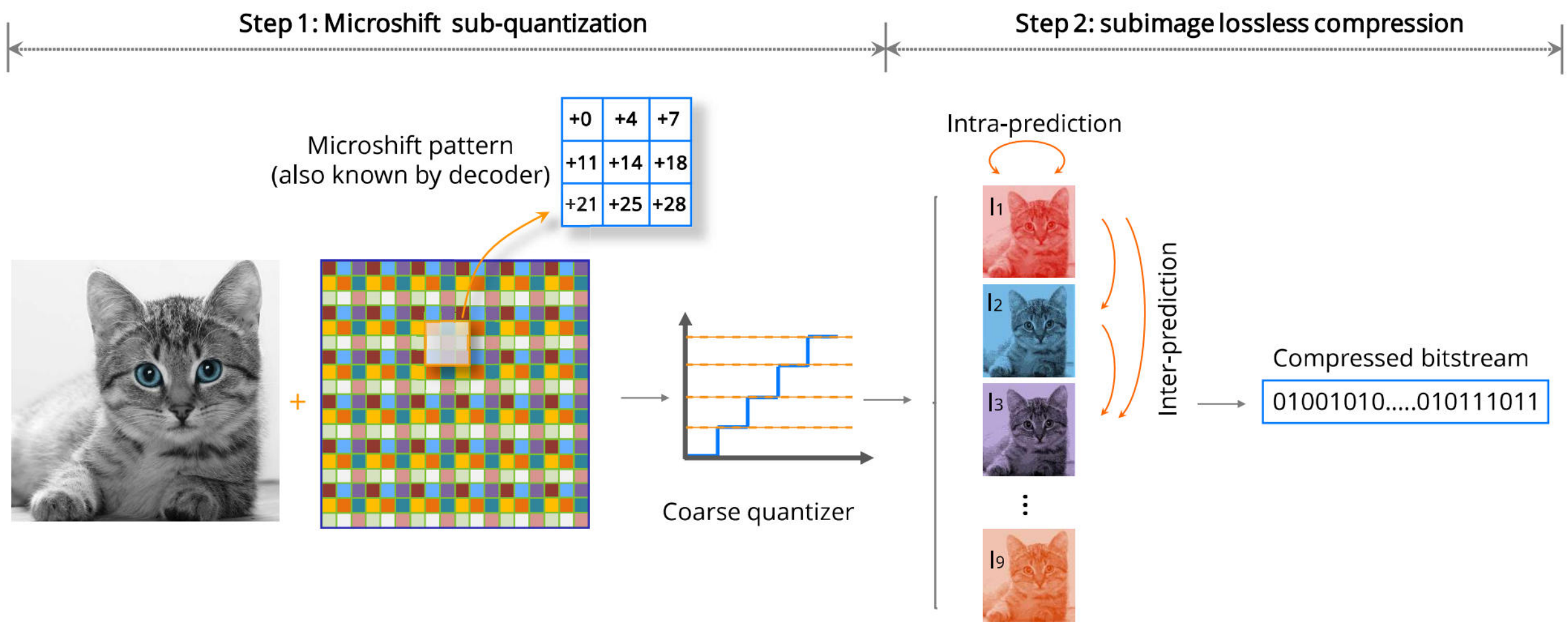

Fig. 1. Compression overview. The first step quantizes the micro-shifted image with $M$ bits. The second step compresses the subimages sequentially.

The remainder of the paper is organized as follows: after reviewing the related work in Sec. II, we propose the *Microshift* algorithm in Sec. III. Both FAST and MRF decompression methods are introduced in Sec. IV. In Sec. V, we extensively evaluate our algorithm. Using the optimal parameter settings obtained from the algorithm evaluation, the hardware implementation architecture is proposed in Sec. VI. Finally, we conclude the paper in Sec. VII.

## II. Related Work

Image compression algorithms can be classified as lossless or lossy. Readers may refer to [1], [2], [12] for a comprehensive review on this topic. Here, we mention the most relevant compression methods, which are later used for comparison with our approach.

Lossless compression can fully recover the image without any information loss. Typical lossless compression algorithms, such as FELICS [13], LOCO-I [14], [15] and CALIC [16], adopt the context-based predictive framework [17], which predicts each pixel based on its context, and then encodes the prediction error with entropy coding techniques. In our compression algorithm, we employ a similar framework to compress the subimages. The major difference is that we predict pixels using a learned predictor to reduce the intra-redundancy in the first subimage; for the subsequent subimages, we propose inter-prediction which uses the information of the subimages that have already been coded.

Typically, lossless compression can only provide a limited compression ratio [12], [18]. Therefore, many transform based lossy compression algorithms have been proposed to greatly improve the compression ratio [19]–[23]. Some works propose hardware implementation architectures for these methods [24]–[26]; however, such dedicated image signal processors (ISPs) are usually costly and power hungry.

In order to achieve low implementation complexity, on-chip compression algorithms have been proposed recently [6]. The block based method proposed in [27] uses quadtree decomposition (QTD) to encode the difference between the current pixel and the brightest pixel in the block. Another approach proposes a 1-bit prediction scheme and uses QTD to encode the prediction error [28]. However, these works are designed for digital pixel sensors (DPSs), while the active pixel sensors (APS) are today's mainstream technology for image sensors due to their higher fill factor and superior image quality [29]. By adopting Morton scan for readout, as proposed in [30], the pixels in the same block can be efficiently averaged, thus achieving adaptive resolution and data compression. In [31], a visual pattern image coding (VPIC) based method is implemented. It compresses an image adaptively according to the patch uniformity, allowing the sensor operate with extremely low power. However, because of their simplicity, these methods do not provide high compression performance. Recently, there have been efforts to integrate compressive sensing techniques into the image sensor design [32]–[34]. Nonetheless, the compressive sensing techniques, though elegant in theory, have to be approximated during real implementation, and the simplification causes significant degradation of the image quality.

Our work is inspired by the PSD algorithm [10], which proposes the basic idea of recovering the image from micro-shifted values. In our work, we greatly improve the compression ratio through a subimage compression step. The problem of dynamic range loss is also addressed. Furthermore, we propose the two methods for decompression, producing significantly better image quality and allowing progressive reconstruction. Lastly, we propose the energy efficient implementation architecture, proving the algorithm highly applicable for low-power WVSNs.

## III. Compression Algorithm

In this section, we introduce our compression algorithm. The first step of the compression is lossy, where the bitdepth is reduced during quantization. In the second step, we further improve the compression ratio by losslessly encoding the subimages using either intra-prediction or inter-prediction.

### A. Microshift based compression

The first step of our compression framework (Fig. 1) builds on the method proposed in [10]. Suppose we use fewer bits (e.g., 3 bits) to represent each pixel. Inferring the original bitdepth from fewer bits is an ill-posed problem. Fortunately, in natural images, neighboring pixels are often correlated. If we quantize the local image patches with shifted quantization levels, we can exploit the spatial redundancy and more accurately estimate the original bitdepth.

Formally, let $i$ be the pixel index of image $I$. Initially, each pixel $I_i$ has an 8-bit bitdepth and takes values in the $[0, 255]$ range. Next, we want to quantize the image using a quantizer whose resolution is $M$-bit ($M < 8$). Therefore, the quantization step is $\Delta = 256/2^M$, and the corresponding quantization levels $\mathcal{L}_k$ are

$$\mathcal{L}_k = k\Delta, k = 0, 1, .., (2^M - 1). \tag{1}$$

Now let us define a microshift pattern with size $N \times N$, which is known by both the encoder and decoder:

$$\mathcal{M}_{pattern} = \begin{bmatrix} \delta_0 & \delta_1 & ... & \delta_{N-1} \\ \delta_N & \delta_{N+1} & .. & \delta_{2N-1} \\ .. & .. & ... & .. \\ \delta_{(N-1)N} & \delta_{(N-1)N+1} & ... & \delta_{N^2-1} \end{bmatrix}, \tag{2}$$

where, $\delta_t$ specifies the corresponding shift, which is:

$$\delta_t = \text{round}(\frac{t}{N^2}\Delta). \tag{3}$$

Then repeat this microshift pattern, so that a microshift array with the same size of the image $I$ can be obtained as follows:

$$\mathcal{M} = \begin{bmatrix} \mathcal{M}_{pattern} & \mathcal{M}_{pattern} & \cdots \\ \mathcal{M}_{pattern} & \mathcal{M}_{pattern} & \cdots \\ \vdots & \vdots & \vdots \\ \mathcal{M}_{pattern} & \mathcal{M}_{pattern} & \cdots \end{bmatrix}. \tag{4}$$

Using the microshift array, we obtain the corresponding shift to each pixel, and then quantize the shifted values using the coarse quantizer with $M$-bit resolution. Finally, we get a sub-quantized micro-shifted image $\widetilde{I}$:

$$\widetilde{I} = \mathcal{Q}(I + \mathcal{M}). \tag{5}$$

Here, $\mathcal{Q}(\cdot)$ denotes the quantization for each pixel using the quantization levels $\mathcal{L}_k, k = 0, 1, .., (2^M - 1)$. The microshift on each pixel is computed through element-wise additions.

*1) Heuristic decompression:* In [10], a heuristic method is proposed to reconstruct the original image $I$ from $\widetilde{I}$. Here, we briefly review it to reveal that the bitdepth information is still preserved in the microshift images. During the proposed compression procedure, we will use this heuristic method for online encoding.

Suppose the pixel $i$ takes the value $\widetilde{I}_i = \mathcal{L}_m$ in the micro-shifted image. Because before the quantization the pixel value is shifted by $\delta_m$, the pixel $i$ is equivalently quantized to $(\mathcal{L}_m - \delta_m)$, with the uncertainty range

$$\mathbb{U}_i = [\mathcal{L}_m - \delta_m, \mathcal{L}_m - \delta_m + \Delta]. \tag{6}$$

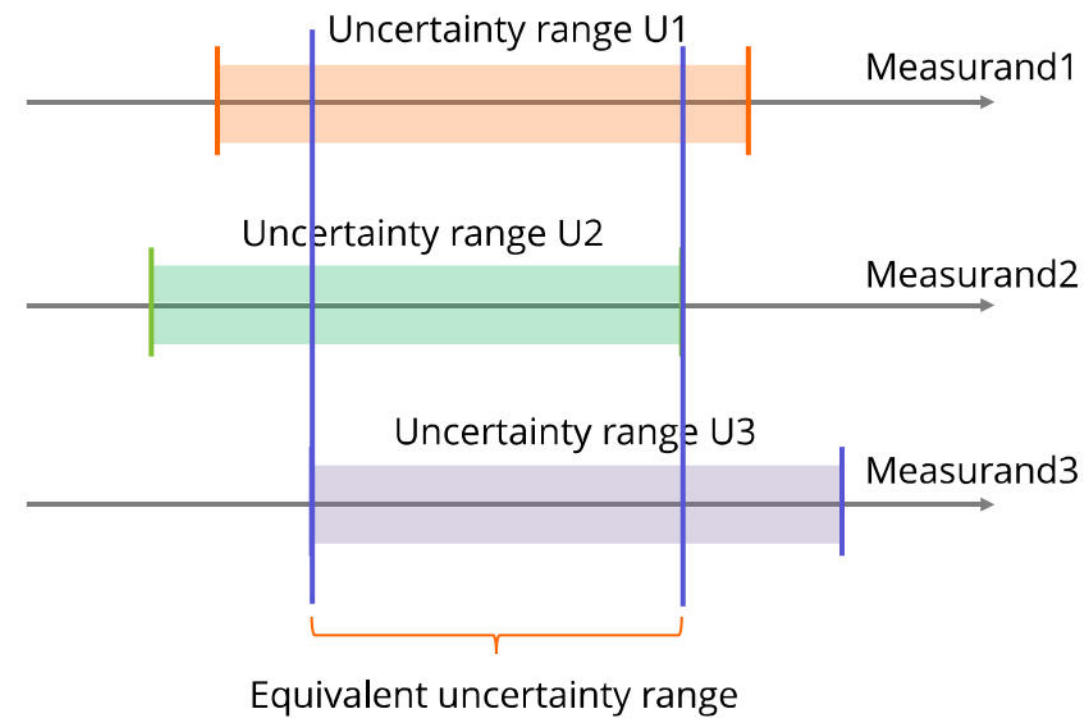

Fig. 2. Pixels that share the same value are quantized by shifted quantizers. The equivalent uncertainty range is the intersection of the uncertainty ranges in the multiple quantizations.

The set of neighborhoods of $i$ within the microshift pattern is denoted as $\mathcal{N}_i$, and the uncertainty ranges of these neighboring pixels $\mathbb{U}_j (j \in \mathcal{N}_i)$ can be obtained similarly. From the observation of $\widetilde{I}_i$ alone, the best estimation of the pixel value is $I_i = \mathcal{L}_m - \delta_m + (1/2)\Delta$, which is the median of the uncertainty range, and the uncertainty is $\pm(1/2)\Delta$. Fortunately, in natural images, neighboring pixels are somewhat correlated. In [10], images are assumed to be piecewise constant, so the pixel $i$ is regarded to be quantized based on all pixels on its neighborhood. Each time, a shifted quantizer is used (Fig. 2), thus the uncertainty range becomes the intersection of the uncertainty ranges of the pixel $i$ and its neighborhood:

$$\mathbb{U}'_i = \bigcap_{j \in \mathcal{N}_i \bigcup \{i\}} \mathbb{U}_j. \tag{7}$$

Albeit simplistic, this heuristic method is effective. However, the assumption that images are piecewise constant is rather strong, making it fail to adapt to edges and textures. Also, the bitdepth can no longer be fully recovered to 8-bit. For example, if we choose a microshift pattern with size $N = 3$ and quantize the micro-shifted images with $M = 3$ bits, we can only recover the image with an uncertainty of $((256/2^M)/N^2) = 3.56$ or equivalent bitdepth of 6.2-bit. In Sec. 4, we will address these issues.

*2) Dynamic range loss:* Another issue is that, after the microshift sub-quantization, 8-bit micro-shifted values may be clamped to the maximum value of 255. As a result, the bright pixels can never be well reconstructed. We refer to this as a dynamic range loss. Fig. 3(b) shows that the sky, which should have been a bright region, is slightly darker after reconstruction. In this work, we propose to overcome this issue by wrapping the micro-shifted value if there is an overflow: if the shifted value is greater than 255, we take the modulo, so Eq. 5 becomes

$$\widetilde{I} = \mathcal{Q}(\text{mod}(I + \mathcal{M}, 256)). \tag{8}$$

Fig. 3(c) shows the modulo micro-shifted image. If the micro-shifted values of bright pixels exceed 255, they will be wrapped to dark pixels. In the highlight regions, there always exist pixels that are shifted by zero and remain bright. Therefore, during the decompression, we can use these pixels to infer the values of dark pixels that are likely to be bright. Fig. 3(d) shows the image decompressed from Fig. 3(c) with negligible dynamic range loss.

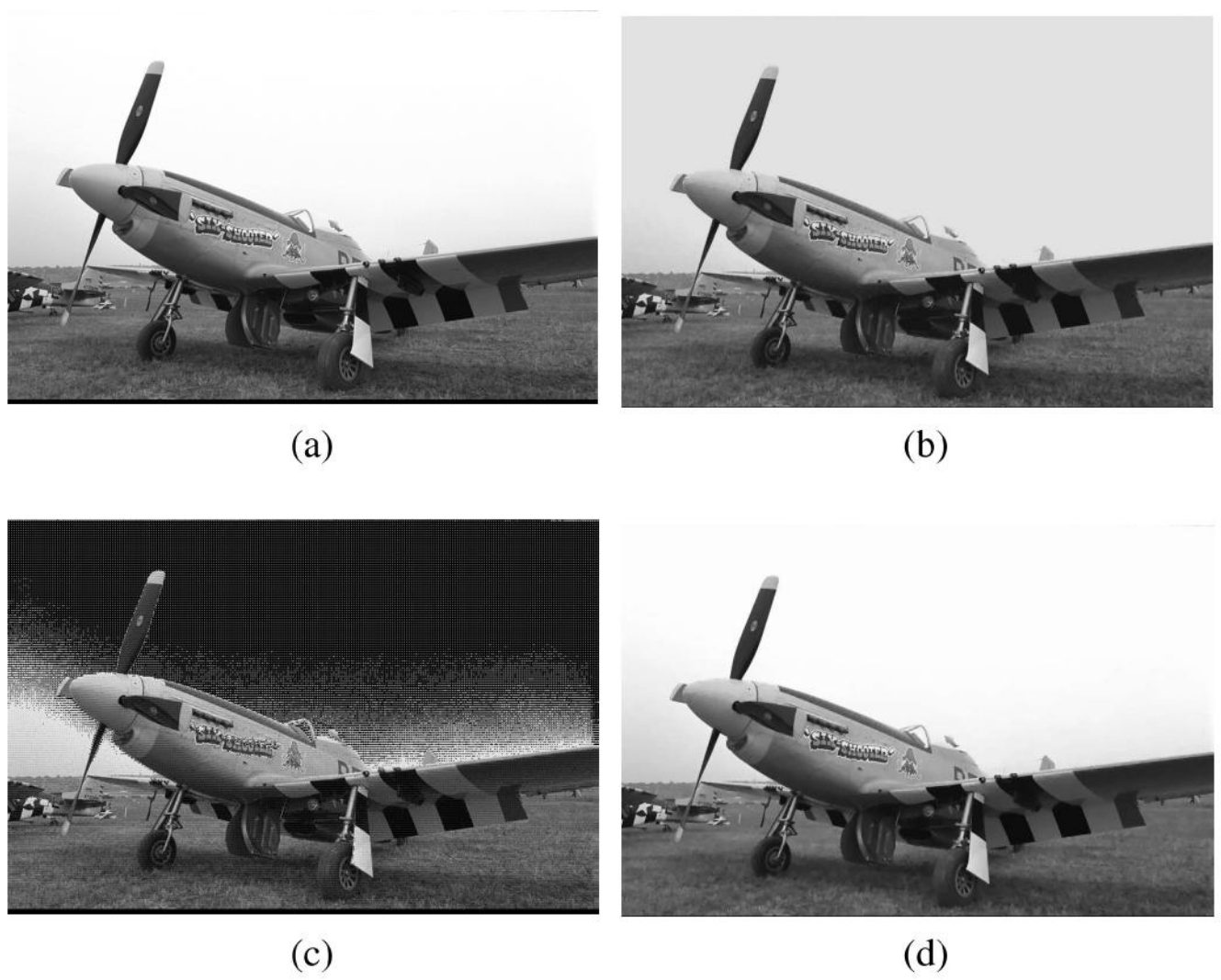

Fig. 3. Dynamic range. (a) Original image. (b) Decompressed image. The bright region cannot be fully recovered due to the dynamic range loss. (c) Modulo microshift image. (d) Reconstructed result from the modulo image.

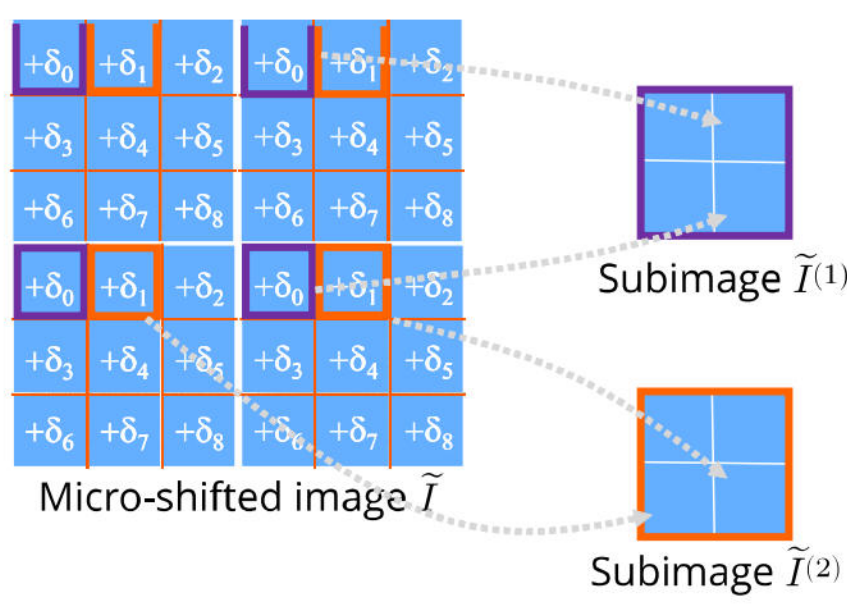

Fig. 4. Each subimage contain pixels that use the same microshift.

### *B. Further compression*

Because the micro-shifted image $\widetilde{I}$ uses an $M$-bit quantizer, while the pixels in the original image $I$ are represented by 8 bits, the compression ratio so far is $CR_1 = 8/M$. Next, we will show that by using a lossless encoding step, the compression ratio can be improved significantly.

*1) Subimage downsampling:* Because of the microshifts, smooth regions become uneven after the quantization, as shown in Fig. 5(a). Encoding the microshift image directly is difficult because of many high frequency components in local regions. However, the image $\widetilde{I}$ can be divided into subimages so that each subimage is more "suitable" for compression. Subimages $\widetilde{I}^{(m)}$ $(m = 1, 2, ..., N^2)$ are formed by downsampling the image $\widetilde{I}$, and they contain pixels that have the same microshift $\delta_m$, as shown in Fig. 4.

The advantage of dividing $\widetilde{I}$ into subimages is two-fold. First, pixels that are quantized by different quantizers are decoupled and pixels in the same subimage share the same quantizer. As a result, large areas of uniform regions are observed in the subimages (Fig. 5(b)), making images more compressible. Second, the downsampling can be regarded as an interlacing technique [35]. Subimages are transmitted in sequence, and the decompression can be progressive without waiting for the transmission to complete. This feature is extremely useful in for low power WVSNs.

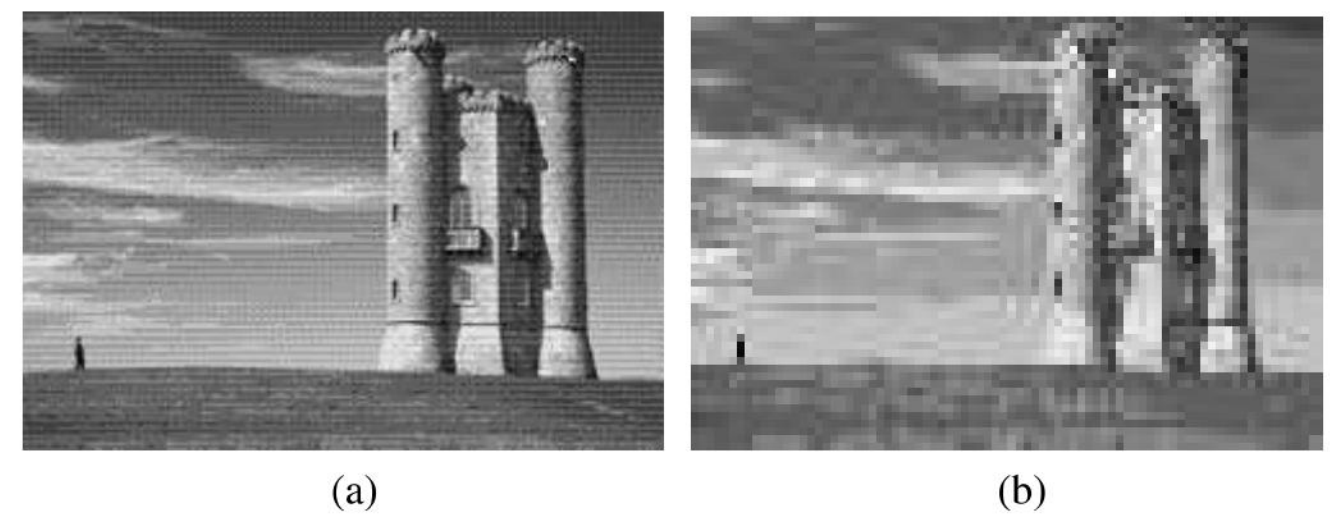

Fig. 5. (a) The microshift image $\widetilde{I}$ using Eq. 5. (b) The subimage $\widetilde{I}^{(j)}$. Note that the dimensions of $\widetilde{I}^{j}$ is actually $1/3$ of the original image.

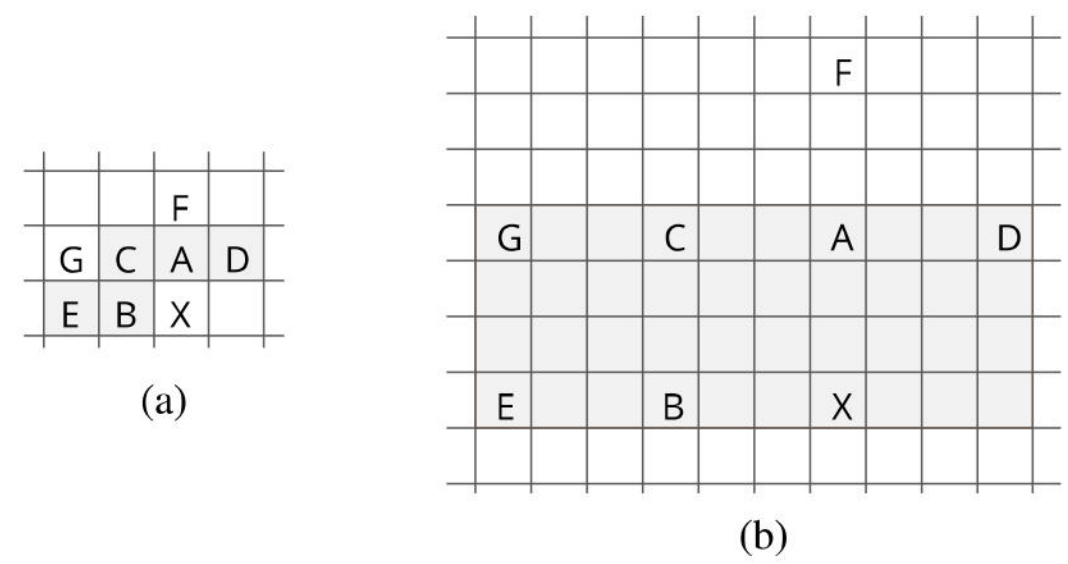

Fig. 6. Prediction template. (a) $A \sim E$ in the subimage $\widetilde{I}^{(1)}$ are used for intra-prediction. (b) the locations of template pixels in the original image $\widetilde{I}$. Rows that are buffered for are denoted in shaded gray.

The first subimage $\widetilde{I}^{(1)}$ is first compressed. Because only the information within $\widetilde{I}^{(1)}$ can be used, we encode the pixel in $\widetilde{I}^{(1)}$ by intra-prediction. Using a different strategy, the subsequent subimages will be compressed using the subimages that have been encoded. We call such a process inter-prediction, and the inference across the subimages can be denoted as

$$\widetilde{I}^{(1)} \rightarrow \widetilde{I}^{(2)},\ \{\widetilde{I}^{(1)}, \widetilde{I}^{(2)}\} \rightarrow \widetilde{I}^{(3)}, ... \tag{9}$$

*2) Intra-prediction:* We compress the first subimage $\widetilde{I}^{(1)}$ through a typical lossless prediction framework. Each pixel of $\widetilde{I}^{(1)}$ is predicted according to its neighboring pixels, then the prediction errors are encoded through entropy coding.

Pixels are raster scanned. The prediction for the current pixel $X$ is based on the causal template shown in Fig. 6(a). We will use the template pixels to estimate $X$. As opposed to the common practice which predicts $X$ directly, we predict the relative difference $X - B$ as suggested in [18].

The size of the prediction template should not be too large because the template pixels are actually not close to $X$ in the original location, as shown in Fig. 6(b). Therefore, we only consider pixels $A \sim F$, whose distances to $X$ are within two pixels in $\widetilde{I}^{(1)}$. Furthermore, because our compression is supposed to run on a single raster scan, the row data are stored in the line buffers. The pixel $F$ has to be excluded for intra-prediction, otherwise we need three more line buffers for accessing $F$, which will bring significant area overhead.

Having the causal template pixels $A \sim E$, we define a texture vector $\mathbf{v}$ whose elements are the differences between these pixels:

$$\mathbf{v} = (A - C \quad C - B \quad D - A \quad B - E). \tag{10}$$

Then each element $\mathbf{v}_i$ can be quantized into the regions of $\{(-T, -T+1), (-T+1, -T+2), ..., (T-1, T)\}$. We then use the quantized texture vector $\widehat{\mathbf{v}} = \{\widehat{\mathbf{v}}_i\}$ to distinguish contexts around $X$. In our work, we choose $T = 2$, so we obtain $(2T+1)^4 = 625$ contexts. In fact, the texture $\widehat{\mathbf{v}}$ and its counterpart $-\widehat{\mathbf{v}}$ can be regarded as equivalent, so we merge these contexts and finally get $(625 - 1)/2 + 1 = 313$ different contexts.

Next, in order to determine the prediction for $X$ according to its context, we propose a learning based predictor instead of a handcrafted explicit function as previous methods [17]. We collect 98 natural images for training, which comprise various categories such as portraits and landscapes. Each image in the training set is also sub-quantized with $M$-bits. We scan all the pixels of the training images and compute the corresponding contexts. For each context, we obtain a histogram of $X - B$. For the pixel whose context index is $l$ ($l \in [1, 313]$), the most probable value in the $l$-th histogram is used to predict the $X - B$, so the prediction $\widehat{\Delta}^l_{X-B}$ can be formulated as

$$\widehat{\Delta}^{(l)}_{X-B} = \text{argmax}\, Histogram_{X-B}(l). \tag{11}$$

Once the predictors are learned, they are stored as a dictionary:

$$\mathcal{D} = [\widehat{\Delta}^{(1)}_{X-B}, \widehat{\Delta}^{(2)}_{X-B}, ..., \widehat{\Delta}^{(313)}_{X-B}], \tag{12}$$

which contains 313 values that are used to achieve the best prediction in different contexts. Then, we can simply predict $X$ using the corresponding dictionary entry:

$$\widehat{X} = B + \mathcal{D}(l). \tag{13}$$

Having the prediction, we can calculate the prediction error to be

$$\epsilon = X - \widehat{X}. \tag{14}$$

Because $X$ is represented by $M$ bits and takes values from $[0, 2^M - 1]$, the prediction error $\epsilon$ has only $2^M$ possibilities: $\epsilon_{possible} = \{-\widehat{X}, 1-\widehat{X}, ..., 2^M-1-\widehat{X}\}$. For each possible $\widehat{X}$, the prediction error can actually be mapped to natural numbers $\{0, 1, ..., 2^M\}$. When $\widehat{X} < 2^{(M-1)}$, which means a positive $\epsilon$ is more probable, the prediction error can be mapped through

$$\epsilon' = \begin{cases} \min(\epsilon - 1, -\epsilon_{possible,min}) + \epsilon, & \text{for } \epsilon > 0 \\ \min(-\epsilon, \epsilon_{possible,max}) + (-\epsilon), & \text{for } \epsilon < 0 \end{cases} \tag{15}$$

Similarly, when $\widehat{X} > 2^{(M-1)}$ and it is more likely to have a negative $\epsilon$, we map the prediction error using

$$\epsilon' = \begin{cases} \min(\epsilon, -\epsilon_{possible,min}) + \epsilon, & \text{for } \epsilon > 0 \\ \min(-\epsilon - 1, \epsilon_{possible,max}) + (-\epsilon), & \text{for } \epsilon > 0 \end{cases} \tag{16}$$

Finally, the mapped prediction residue $\epsilon'$ will be encoded using entropy coding techniques. In this work, we choose Golomb codes because they are friendly to hardware implementation [36].

The limitation of using Golomb codes alone is that at least 1 bpp (bit per pixel) is needed even if the prediction error is zero. Note that in the sub-image $\widetilde{I}^{(1)}$, there are large areas of uniform regions due to sub-quantization. Therefore, we use adaptive runlength encoding [14] specifically for these regions.

**Algorithm 1** *Microshift* compression

**Input:** The raw image $I$ with 8-bit values
**Output:** The compressed bitstream

**for** each pixel $X$ in $I$ **do**
  Microshift quantization: $\widetilde{X} = \mathcal{Q}(\text{mod}(X + \mathcal{M}, 256))$
  Compute the texture vector $\mathbf{v}$
  Quantize the texture vector: $\widehat{\mathbf{v}} = \mathcal{Q}(\mathbf{v})$
  **if** $\mathbf{v}$ is uniform **then**
    Runlength encode
  **else**
    **if** $\widetilde{X} \in \widetilde{I}^{(1)}$ **then**
      Determine the context index $l$
      Calculate the prediction: $\widehat{X} = B + \mathcal{D}(l)$
      $\epsilon_{prediction} \leftarrow X - \widehat{X}$
    **else**
      Access corresponding pixels: $\widetilde{X}_1, \widetilde{X}_2, ..., \widetilde{X}_{j-1}$
      $\widehat{X}_j \leftarrow \mathcal{Q}[Dec(\widetilde{X}_1, \widetilde{X}_2, ..., \widetilde{X}_{j-1}) + \delta_j]$
      $\epsilon_{prediction} \leftarrow X_j - \widehat{X}_j$
    **end if**
    $\epsilon_{mapped} \leftarrow \text{map}(\epsilon_{prediction})$
    $\text{GolombEncode}(\epsilon_{mapped})$
  **end if**
**end for**

*3) Inter-prediction:* For the subsequent subimages $\widetilde{I}^{(2)} \sim \widetilde{I}^{(9)}$, we propose inter-prediction for compression, which yields more efficient results than intra-prediction for these subimages.

Suppose we want to compress the $j$th subimage $\widetilde{I}^{(j)}$ ($j \geq 2$). Instead of using the intra-predictor, which uses the information within the same image, we find that the pixels in $\widetilde{I}^{(j)}$ can be efficiently predicted using $\widetilde{I}^{(1)} \sim \widetilde{I}^{(j-1)}$, subimages that have been encoded during the sequential compression. This is because the corresponding pixels in $\widetilde{I}^{(1)} \sim \widetilde{I}^{(j-1)}$ are actually closer to the current pixel $X_j$ than its neighboring pixels within $\widetilde{I}^{(j)}$, as shown in Fig. 6(b). Denote the corresponding pixels in $\widetilde{I}^{(1)} \sim \widetilde{I}^{(j-1)}$ to be $X_1, X_2, ..., X_{j-1}$, then the original value at the location $X_j$ can be estimated by running the heuristic decompression, and $X_j$ in the $j$th microshift image can be predicted as

$$\widehat{X}_j = \mathcal{Q}[Dec(\widetilde{X}_1, \widetilde{X}_2, ..., \widetilde{X}_{j-1}) + \delta_j]. \tag{17}$$

Next, the prediction error can be encoded through Eq. 14-16 similar to that used in the intra-prediction.

Furthermore, similar to intra-prediction, using inter-prediction alone only achieves 1 bpp at best. To improve this, the compression will also go into runlength mode if the current context is uniform, just as in intra-prediction. Algorithm 1 summarizes the overall *Microshift* compression.

*4) Extension for color images:* Our algorithm can also compress color images. Contrary to common practice which uses color transformation to decorrlate color channels [18], [25], our algorithm processes the RGB channels individually. We adopt such color handling technique because the algorithm can directly process the Bayer pattern (BGGR pattern) on the image sensor, making the method hardware friendly and suitable for the circuitry front-end implementation.

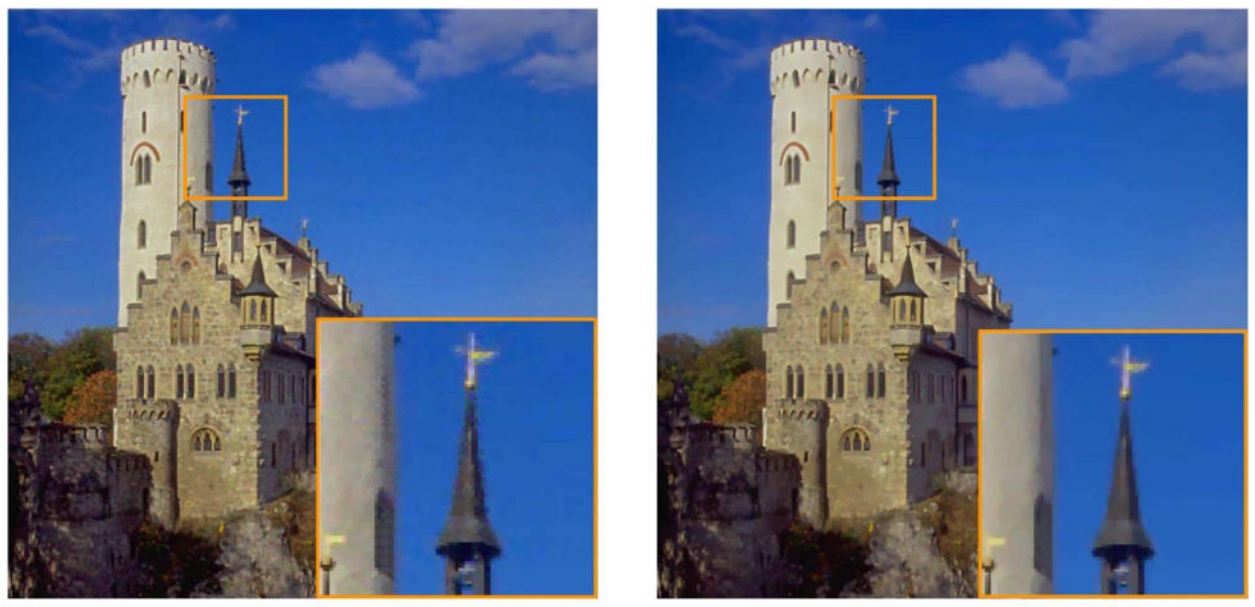

(a) (b)

Fig. 7. Decompression result. (a) Decompressed image using FAST. (b) Decompressed image using the MRF model shows crisp boundaries.

## IV. Decompression algorithm

The *Microshift* compression in Sec. III consists of two steps, so the decompression can be accomplished by simply reversion the operation. Since the subimages are compressed losslessly, they can be fully recovered using the causal pixels through intra- or inter-prediction. The lossless decompression follows the same steps of Algorithm 1 in reverse order.

Having the subimages $\widetilde{I}^{(1)} \sim \widetilde{I}^{(9)}$, we can combine them and obtain the micro-shifted image $\widetilde{I}$, whose bitdepth is $M$-bit. In this section, we propose two methods–FAST and MRF model–to reconstruct the original image $I$. Progressive decompression extensions will be discussed at the end of this section.

### A. FAST decompression

Though the least significant bits are lost during the sub-quantization, they can still be inferred from the bitdepth of the neighboring pixels using the heuristic decompression in III-A. This method operates extremely fast and produces decent image quality. However, there are some problematic artifacts. First, edges in the decompressed image may have a sawtooth appearance. Second, the decompression output looks a bit noisy because each pixel is inferred just from its neighborhood and the decompression of each pixel is independent. Third, quantization artifacts are noticeable on the smooth regions because the bitdepth is not fully recovered to 8 bits.

To alleviate these artifacts, one approach is to employ an edge-preserving image filter during post-processing, so that a larger context can be utilized for the decompression for each pixel. We use the fast weighted least square (WLS) filter from [11] because it is an efficient global smoother ($\mathcal{O}(n)$) and will not induce "Halo" artifacts [37]. We iteratively apply the filter 8 times and the image quality improves considerably. Fig. 7(a) shows the decompressed result. Because the filter runs efficiently (0.7s to process a 512x512 image[1]), we refer to this decompression method as FAST.

### B. MRF decompression

Using an image smoother for post-processing is not enough. In Fig. 7(a), the artifacts, such as the sawtooth edges, are still clearly visible, which severely affect the perceptual quality. The reason for such artifacts is that in the heuristic model local patches are assumed to be piece-wise constant [10]; therefore, the decompression cannot adapt to edges or textures very well. In order to overcome the limitation, we propose a Markov random field (MRF) model [38], [39] for decompression. In this model images are assumed to be piece-wise smooth rather than piece-wise constant.

Still, we denote the sub-quantized microshift image to be $\widetilde{I}$, which can be decompressed losslessly from the bitstream. The pixels $\mathbf{z} = (z_1, z_2, ..., z_n)$ in $\widetilde{I}$ are the observations and each $z_i$ takes its value from the quantization levels $\mathcal{L}_k$. Our aim is to find the most probable pixel values $\mathbf{x} = (x_1, x_2, ..., x_n)$ in the original image $I$ according to the observations. These values can be inferred from the maximum a posterior (MAP) perspective:

$$\widehat{x}_{1...n} = arg \max_{x_{1...n}} Pr(x_{1...n}|z_{1...n}), \tag{18}$$

where $Pr(x_{1...n}|z_{1...n})$ is the posterior probability given the observations. Using the Bayesian rule and transformation to the log domain, we have

$$\begin{aligned}\widehat{x}_{1...n} &= \arg\max_{x_{1...n}} Pr(z_{1...n}|x_{1...n}) \cdot Pr(x_{1...n}) \\ &= \arg\min_{x_{1...n}} [-\log Pr(z_{1...n}|x_{1...n}) - \log Pr(x_{1...n})],\end{aligned} \tag{19}$$

where, $Pr(z_{1...n}|x_{1...n})$ is the likelihood of the observations $\mathbf{z}$ given the pixel values $\mathbf{x}$, and $Pr(x_{1...n})$ is the prior probability of the pixel values $\mathbf{x}$. Furthermore, Eq. 19 can be written as:

$$\widehat{x}_{1...n} = arg \min_{x_{1...n}} [\sum_{i=1}^{n} U_i(x_i, z_i) + \gamma \sum_{(p,q)\in\mathcal{C}} P_{pq}(x_p, x_q)], \tag{20}$$

where, $U_i(x_i, z_i)$ is the data term, which represents the cost of choosing $x_i$ for estimation when observing $z_i$. It is an unary function of $x_i$:

$$U_i(x_i, z_i) = -\log Pr(z_i|x_i). \tag{21}$$

In Eq. 20, the notation $\mathcal{C}$ represents an 8-connected clique and the pairwise term $P_{pq}(x_p, x_q)$ encodes the prior of the original pixel values, which penalizes the smoothness of the optimization output. $\gamma$ is the coefficient balancing the two terms. Next, we will model the data term and smoothness term individually.

*1) Data term:* According to Eq. 21, to formulate the data term, we need to model the likelihood of the observations $\mathbf{z}$. Let us recall the generative model in Eq. 5. The image $\widetilde{I}$ is formed by quantizing the micro-shifted image. Specifically, the generation of the observations $\mathbf{z_i}$ can be formulated as:

$$\widetilde{x}_i = x_i + \delta_i + \xi_i \tag{22}$$

$$z_i = \mathcal{Q}(\widetilde{x}_i), \tag{23}$$

where, $\delta_i$ is the corresponding microshift, $\xi_i$ is the quantization noise introduced during the sub-quantization, and $\widetilde{x}_i$ denotes the micro-shifted value before sub-quantization. We approximate the quantization noise by a normal distribution

[1]Measured on a single thread of Intel i5 3.2GHz CPU

$\xi_i \sim \mathcal{N}(0, \sigma^2)$, where $\sigma^2$ is the variance of the distribution. Because $\xi_i = \widetilde{x}_i - x_i - \delta_i$, we have

$$Pr(\widetilde{x}_i|x_i) = \frac{1}{\sqrt{2\pi\sigma^2}} \exp[-\frac{(\widetilde{x}_i - x_i - \delta_i)^2}{2\sigma^2}]. \quad (24)$$

As Eq. 23 shows, the observation $z_i$ is sub-quantized from $\widetilde{x}_i$. Because $z_i$ is generated from $\widetilde{x}_i$ which ranges in $[z_i, z_i + \Delta)$, the probability of $z_i$ given $x_i$ should be an integral of $Pr(\widetilde{x}_i|x_i)$ for all the possible $\widetilde{x}_i$, which is

$$\begin{aligned} Pr(z_i|x_i) &= \int_{z_i}^{z_i+\Delta} Pr(\widetilde{x}_i|x_i) \cdot \frac{1}{\Delta} d\widetilde{x}_i \\ &= \frac{1}{2\Delta}[\mathrm{erf}(\frac{z_i - x_i - \delta_i + \Delta}{\sqrt{2\sigma^2}}) \\ &\quad - \mathrm{erf}(\frac{z_i - x_i - \delta_i}{\sqrt{2\sigma^2}})], \end{aligned} \quad (25)$$

where $\mathrm{erf}(\cdot)$ is the error function. Equation 25 is similar to the model in [40]. The difference is that in [40] the bit-depth is expanded for natural images, whereas we model the likelihood for micro-shifted images and aim to improve the image quality for decompression.

*2) Smoothness term:* The smoothness term $Pr(x_{1...n})$ is a pairwise function that measures the interaction of each pixel pair. The smoothness cost is defined to be

$$P_{pq}(x_p, x_q) = \lambda_{(p,q)} \mu_{(z_p,z_q)} \nu_{(\delta_p,\delta_q)} |x_p - x_q|, \quad (26)$$

where $p$ and $q$ represent the locations of neighboring pixels, and the smoothness cost encourages coherent regions. In Eq. 26, $\lambda_{(p,q)}$, $\mu_{(z_p,z_q)}$ and $\nu_{(\delta_p,\delta_q)}$ are adaptive weights that control the penalization strength. Specifically, $\lambda_{(p,q)}$ is inversely proportional to the pixel distance:

$$\lambda_{(p,q)} = 1/\,\mathrm{dist}(p, q), \quad (27)$$

where $\mathrm{dist}(\cdot)$ represents the Euclidean distance between pixel $p$ and $q$. The higher the spatial distance between $p$ and $q$, the smaller their interaction. Besides, $\mu_{(z_p,z_q)}$ is a function of the intensity difference:

$$\mu_{(z_p,z_q)} = \begin{cases} 1, & \text{if } |(z_p - \delta_p) - (z_q - \delta_q)| < T \\ 0, & \text{otherwise} \end{cases}, \quad (28)$$

which is a binary number according to the intensity similarity relative to the threshold $T$. There is an interaction between $p$ and $q$ only if their intensity is close enough. The weights $\lambda_{(p,q)}$ and $\mu_{(z_p,z_q)}$ together can be regarded as a bilateral coefficient, so the regularization will be edge-aware.

Besides, in this work, even the pixels which are originally intensity-close, may appear different due to the distinct microshifts (Fig. 5(b)). Therefore, they may be mis-classified to different contexts during the edge-aware optimization. In order to compensate such an issue, we propose $\nu_{(\delta_p,\delta_q)}$ in Eq. 26, which is defined heuristically using a logistic function:

$$\nu_{(\delta_p,\delta_q)} = \frac{1}{1 + \exp(-\alpha|\delta_p - \delta_q|)}. \quad (29)$$

where $\alpha$ is a positive coefficient. The larger the microshift difference is, the larger the compensation weight $\nu$ will be.

*3) Decompress the modulo image:* In order to reconstruct the modulo microshift image that is compressed through Eq. 8, we fine-tune the MRF model (Eq.20) to be:

$$\begin{aligned} \min_{x_{1...n}, \rho_{1...n}} \quad & [\sum_{i=1}^{n} U_i(x_i - \rho_i \cdot 256, z_i) \\ & + \gamma \sum_{(p,q)\in\mathcal{C}} P_{pq}(x_p, x_q)] \\ st. \quad & \rho_i \in \{0, 1\} \end{aligned} \quad (30)$$

where $\rho_i$ is a binary variable indicating whether or not the micro-shifted pixel incurs overflow and has been wrapped to dark. Thanks to the smoothness prior, the modulo pixels can be correctly reconstructed by considering the interactions of adjacent pixels.

*4) Inference:* We solve the optimization problem through the graph cut method which reduces it to a max-flow problem [41], [42]. In order to speed up the solver, we initialize the solution using the result of heuristic decompression. Fig. 7(b) shows the image obtained using the MRF method. Compared to Fig. 7(a), it can be seen that the edges can be reconstructed sharply without sawtooth artifacts because the local image patches are no longer assumed to be constant. The false contours in the sky are less noticeable because the quantization errors are dispersed globally.

### *C. Progressive decompression*

As the data is received, the subimages $\widetilde{I}^{(1)} \sim \widetilde{I}^{(9)}$ are sequentially decompressed. The progressive decompression using the FAST method is simple. The pixels corresponding to the received subimages will be decompressed locally using the template pixels. The remaining locations are then interpolated bilinearly. The progressive decompression result for Lena using the FAST method is shown in Fig. 8(a). Initially, blocking artifacts are clearly observed; as more data is received, both the spatial and bitdepth resolution increase, and the image quality progressively improves.

For the progressive MRF method, the pixels corresponding to the unreceived subimages will be assigned a zero data term. That is, only the smoothness penalty determines the values at these locations. Besides, we set $\gamma$ to grow linearly during the progressive reconstruction. This is because as more subimages are received, more terms are added into the data term and we need to increase the weight of the smoothness accordingly.

Fig. 8(b) shows the quantitative result for the quality improvement using different decompression methods. All the methods steadily improve the PSNR during the progressive decompression. By using an edge preserving filter for post-processing, FAST is consistently better than the heuristic method. On the other hand, MRF decompression is not comparable to these two methods when the received bitstream is too short. However, when the complete bitstream is received, the MRF method improves the PSNR by about 1.5 dB, which is a significant improvement for image quality.

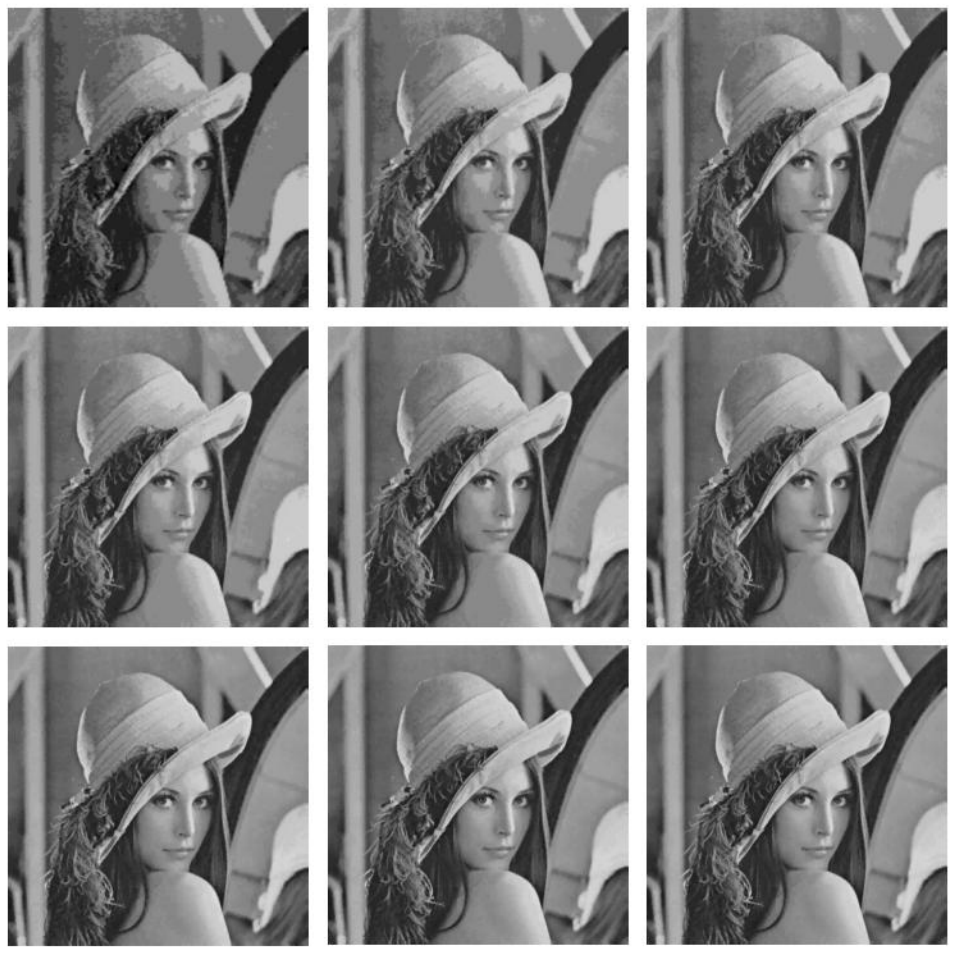

(a)

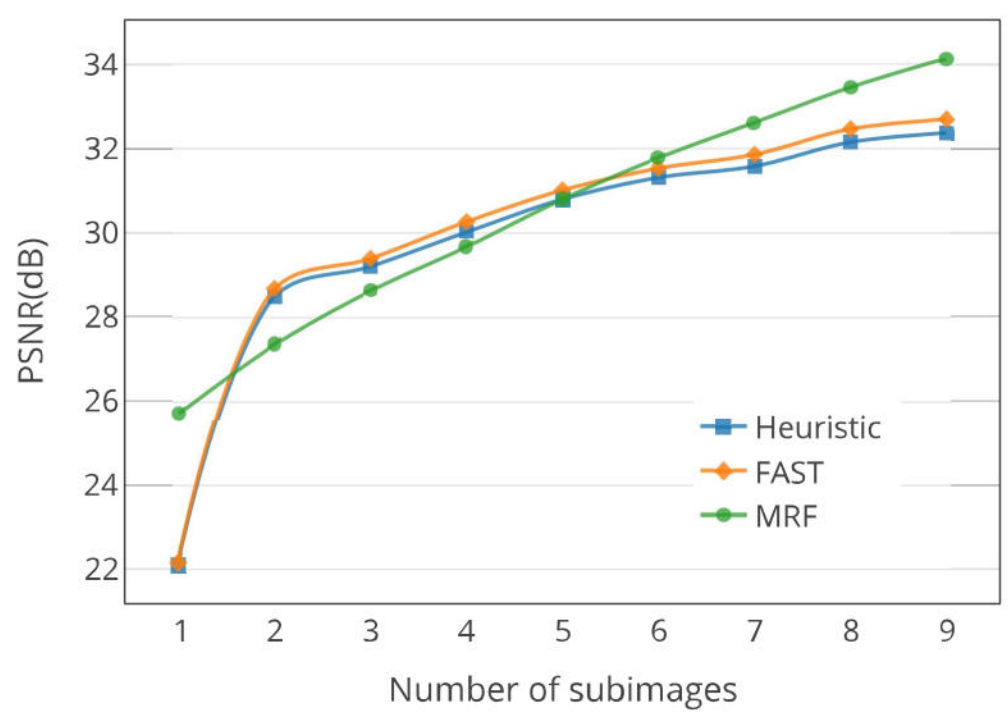

(b)

Fig. 8. Progressive decompression. *Microshift* uses the parameters $N = 3$, $M = 3$ for compression. (a) With longer bitstream of subimages received, the decompressed quality gradually increases. (b) The PSNR increases as more subimages are used for decompression.

## V. Algorithm Experiments

We perform tests on 25 standard images collected from the USC-SIPI dataset[1] and the Kodak dataset[2]. In order to demonstrate the effectiveness of each part of our method, we perform tests for compression and decompression individually. Finally, we compare our method with previous approaches.

### A. Test for Microshift compression

There are two parameters in the *Microshift*: the block size $N$ and the quantization resolution $M$. In order to determine the optimal parameter setting for the following tests, we first investigate how these parameters affect the compression performance. The result is shown in Table I. To keep succinct, experimental results on four test images are shown in the table. Heuristic decompression is used for evaluating the decompression quality.

The compression ratio of the microshift sub-quantization, $CR_1$, only depends on the quantization resolution $M$. A small $M$ leads to a larger $CR_1$. The compression ratio of the subimage compression, $CR_2$, is not sensitive to the increase of $M$. On the other hand, the PSNR increases for a larger $M$ because more bitdepth information is preserved during the quantization. As a result, $M = 3$ is a proper quantization resolution, achieving a good tradeoff between compression ratio and decompression quality.

[1]USC-SIPI dataset: http://sipi.usc.edu/database/

[2]Kodak dataset: http://r0k.us/graphics/kodak/

TABLE I
Influence of different block sizes and bitdepths

| images | block size N | bitdepth M | CR1 | CR2 | PSNR (dB) | BPP (bit) |
|---|---|---|---|---|---|---|
| Lena | 3 | 2 | 4.000 | 1.974 | 29.483 | 1.013 |
| | | 3 | 2.667 | 2.182 | 32.393 | 1.375 |
| | | 4 | 2.000 | 2.036 | 35.430 | 1.965 |
| | 4 | 2 | 4.000 | 1.825 | 27.533 | 1.096 |
| | | 3 | 2.667 | 2.024 | 30.924 | 1.483 |
| | | 4 | 2.000 | 1.861 | 34.504 | 2.149 |
| pepper | 3 | 2 | 4.000 | 2.050 | 29.870 | 0.976 |
| | | 3 | 2.667 | 2.240 | 33.028 | 1.340 |
| | | 4 | 2.000 | 2.111 | 35.703 | 1.895 |
| | 4 | 2 | 4.000 | 1.877 | 28.196 | 1.066 |
| | | 3 | 2.667 | 2.069 | 31.642 | 1.450 |
| | | 4 | 2.000 | 1.926 | 34.823 | 2.077 |
| airplane | 3 | 2 | 4.000 | 2.860 | 25.113 | 0.699 |
| | | 3 | 2.667 | 2.557 | 33.066 | 1.173 |
| | | 4 | 2.000 | 2.179 | 36.735 | 1.836 |
| | 4 | 2 | 4.000 | 2.612 | 23.843 | 0.766 |
| | | 3 | 2.667 | 2.328 | 31.754 | 1.289 |
| | | 4 | 2.000 | 1.946 | 35.791 | 2.055 |
| yacht | 3 | 2 | 4.000 | 1.942 | 27.943 | 1.030 |
| | | 3 | 2.667 | 2.118 | 31.409 | 1.417 |
| | | 4 | 2.000 | 1.941 | 35.552 | 2.061 |
| | 4 | 2 | 4.000 | 1.766 | 25.964 | 1.132 |
| | | 3 | 2.667 | 1.918 | 29.975 | 1.564 |
| | | 4 | 2.000 | 1.717 | 34.484 | 2.330 |

When using a larger block size $N$, $CR_2$ decreases because more subimages will be compressed sequentially, and the run-length becomes shorter due to a smaller subimage resolution. Also, the decompressed image quality becomes worse because larger neighborhoods are used for decompression but some of those pixels are too far from the current pixel. Therefore, $N = 3$ is an optimal block size.

In the following tests and the hardware implementation, we choose $M = 3$ and $N = 3$ as default.

### B. Test for further compression

In the subimage compression, we propose a learning-based intra-predictor to compress the first subimage. We compare our intra-predictor with three other commonly used predictors [17]: GAP, MED, GED. Table II shows the entropy of the prediction residues on 25 images. A lower entropy indicates a better prediction performance. We can see that MED and GED have similar performance, while our learned predictor is significantly better than both of them. Although slightly inferior than GAP, our predictor is tailored not to use the pixel $F$ (Fig. 6), and saves three line buffers in the implementation. Also, no parameter is needed. In all, the proposed predictor is efficient from both the algorithm and hardware perspectives.

Next, we compare the effect of using inter-prediction and intra-predictor in compressing the subsequent subimages. For each test image, the average entropy of prediction residues for the subsequent subimages $\widetilde{I}^{(2)} \sim \widetilde{I}^{(9)}$ is calculated. The result in Fig. 9 shows that for all the test images, the inter-predictor produces significantly lower prediction entropy, which demonstrates the effectiveness of the inter-predictor when compressing the subsequent subimages.

TABLE II
COMPARISON OF ENTROPY OF INTRA-PREDICTION ERROR. LOWER ENTROPY DENOTES BETTER PREDICTION PERFORMANCE.

| images | GAP | MED | GED | LEARN |
|---|---|---|---|---|
| boats | 0.3029 | 0.3285 | 0.3455 | 0.3222 |
| baboon | 0.7409 | 0.7670 | 0.7400 | 0.7475 |
| barbara | 0.5025 | 0.5304 | 0.5636 | 0.5250 |
| flower | 0.2915 | 0.3113 | 0.3353 | 0.3013 |
| flowers | 0.4565 | 0.4843 | 0.5066 | 0.4678 |
| girl | 0.3376 | 0.3688 | 0.3932 | 0.3470 |
| goldhill | 0.4051 | 0.4366 | 0.4322 | 0.4160 |
| lenna | 0.3766 | 0.4043 | 0.4345 | 0.3793 |
| man | 0.4189 | 0.4456 | 0.4499 | 0.4212 |
| pens | 0.3908 | 0.4120 | 0.4044 | 0.4038 |
| pepper | 0.4304 | 0.4753 | 0.4315 | 0.4168 |
| sailboat | 0.4998 | 0.5348 | 0.5182 | 0.5012 |
| tiffany | 0.3162 | 0.3480 | 0.3341 | 0.3183 |
| yacht | 0.3395 | 0.3570 | 0.3952 | 0.3791 |
| Lichtenstein | 0.3436 | 0.3665 | 0.3757 | 0.3613 |
| airplane | 0.3025 | 0.3331 | 0.3313 | 0.3131 |
| cameraman | 0.2525 | 0.2657 | 0.2969 | 0.2690 |
| kodim05 | 0.5676 | 0.5917 | 0.6019 | 0.5786 |
| kodim09 | 0.3364 | 0.3597 | 0.3454 | 0.3493 |
| kodim14 | 0.4898 | 0.5115 | 0.5101 | 0.5094 |
| kodim15 | 0.3232 | 0.3444 | 0.3494 | 0.3295 |
| kodim20 | 0.2806 | 0.3042 | 0.2865 | 0.2849 |
| kodim21 | 0.4510 | 0.4703 | 0.4575 | 0.4637 |
| kodim23 | 0.2736 | 0.2950 | 0.2884 | 0.2719 |
| milkdrop | 0.2201 | 0.2365 | 0.2459 | 0.2330 |
| ***Average*** | **0.3860** | **0.4113** | **0.4149** | **0.3964** |

### C. Overall performance

Finally, we comprehensively compare our method with other on-chip compression algorithms: the PSD algorithm [10], block based compression [27], predictive boundary [28], VPIC [31], block-based compressive sensing [33] and DCT based compression [20]. Because some of the works do not publish their source code, we reproduce the results based on their paper. For block-based compression, we adopt the Hilbert scan for quadtree decomposition, as suggested in the work of predictive boundary [28], so the performance differs slightly from the claimed figure in the original paper. For the compressive sensing, incoherent measurements are acquired independently in each $8 \times 8$ block, which is the common practice in the compressive sensing imagers because the noiselet transform can only be easily implemented in blocks. $\mathcal{L}_1$-magic library [43] is used to recover the compressive sampled image in each block. Finally, in the DCT compression, only runlength encoding with no Huffman encoding is used to encode the transform coefficients.

In this work, we propose different decompression methods, and their performances are evaluated separately. Furthermore, to make a thorough discussion, we also encode the prediction residue using adaptive arithmetic encoding (we still use FAST for decompression, so the decompressed image quality is the same as Golomb encoding), and include the result (denoted as Microshift-arithmetic) in our tests.

Since some works are designed to compress square images, we crop the test images so that their aspect ratios are 1:1, and scale them to the same resolution $512 \times 512$. For fair comparison, we tune the PSNR of different algorithms to around 32 dB, and compare their bit per pixel (BPP) values. For the predictive boundary method, since its decompressed image quality is limited, we tune its BPP value to 1.24 in order to closely match our work.

Table III shows the comparison results. The Microshift algorithm with FAST decompression compresses the image with a much lower BPP than the PSD compression, block-based compression, VPIC, and the block-based compressive sensing (CS). The block-based CS provides limited compression capability partly due to the blocking artifact. Predictive boundary method gives a much lower image quality when maintaining a similar BPP to our work. When tuning the quality level to 25 (the highest quality level is 100), DCT compression gives a compressed image quality similar to Microshift-FAST, but our method can give a lower BPP result. Thus, our method is even more effective than the DCT method for compression. On the other hand, the Microshift-MRF increases the PSNR by about 1 dB on average compared to the FAST decompression, which is a significant improvement in terms of image quality. In Table III, we also include structural similairty (SSIM) index [44], which is a commonly used metric for measuring the perceptual image quality. The Microshift-MRF still shows better decompressed image quality than Microshift-FAST in terms of SSIM, while both of them give much lower BPP than the methods which provide similar perceptual quality. Furthermore, by using arithmetic codes instead of Golomb codes, Microshift-arithmetic can further reduce the BPP by 0.1368. When the coding complexity is not an issue, arithmetic coding can be a good alternative in our method.

## VI. HARDWARE IMPLEMENTATION

In Sec. III, we propose the *Microshift* through a co-design methodology, which considers both the algorithm and hardware efficiency. In Sec. V we validate the performance of our compression algorithm, and in this section, we will introduce the hardware implementation.

### A. Overview of the hardware implementation

The architecture of the hardware implementation is illustrated in Fig. 10. Pixels are read out in a raster scan manner. Then, they will go through a microshift quantization as Eq. 8, and each of the shifted pixels is represented by 3 bits. The quantization makes the following blocks power efficient because they only process the 3-bit values.

Then, these sub-quantized microshifted values are fed into three $W$-stage shifters (image size is $H \times W$), which are connected together in series. These $W$-stage shifters serve as line buffers during the raster scan, which store the image data of the previous lines. The output of each shifter serves as an input to the following 10-stage shift registers, which store the neighborhood of the pixel to be processed. In every clock cycle the quantized image data (3 bit) is read into the first row of the $W$-stage shifters and the data of these shifters are shifted into the corresponding row of the $2\times10$ registers. In this way, the $2\times10$ kernel block scans the entire image and the template pixels A$\sim$E can be accessed for the intra-/inter-prediction during the subimage compression.

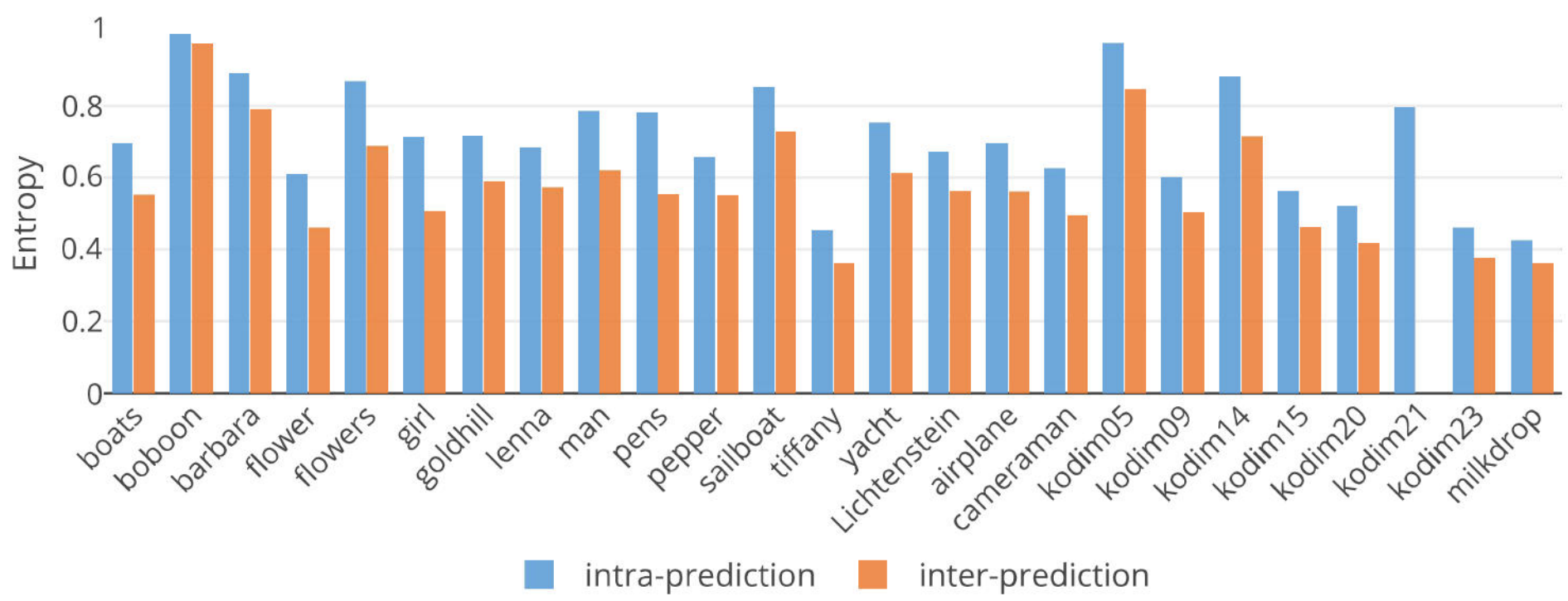

Fig. 9. Entropy for intra-prediction and inter-prediction when encoding the subsequent subimages. Lower entropy indicates better compression performance.

TABLE III
COMPARISON OF COMPRESSION PERFORMANCE

| images | Microshift - FAST | | Microshift – MRF | | PSD [13] | | block-based compression [27] | | predictive boundary [28] | | VPIC [30] | | block-based compressive sensing [33] | | DCT [20] | |
|---|---|---|---|---|---|---|---|---|---|---|---|---|---|---|---|---|
| | **PSNR/SSIM** | **BPP** | **PSNR/SSIM** | **BPP** | **PSNR/SSIM** | **BPP** | **PSNR/SSIM** | **BPP** | **PSNR/SSIM** | **BPP** | **PSNR/SSIM** | **BPP** | **PSNR/SSIM** | **BPP** | **PSNR/SSIM** | **BPP** |
| boats | 32.23/0.9191 | 1.1406 | 33.46/0.9246 | 1.1406 | 32.17/0.9118 | 3.0000 | 31.54/0.9307 | 2.0369 | 25.47/0.6928 | 1.2405 | 31.03/0.9430 | 2.7662 | 32.33/0.9345 | 6.0000 | 33.65/0.9213 | 1.6390 |
| baboon | 28.58/0.8537 | 1.6831 | 29.18/0.8588 | 1.6831 | 28.56/0.8544 | 3.0000 | 28.62/0.9122 | 3.3716 | 22.40/0.6747 | 1.2290 | 26.85/0.8635 | 3.4353 | 28.26/0.8917 | 6.0000 | 27.47/0.8341 | 2.7776 |
| barbara | 30.33/0.8962 | 1.4382 | 31.06/0.8984 | 1.4382 | 30.29/0.8904 | 3.0000 | 30.34/0.9270 | 2.5199 | 24.22/0.6622 | 1.2403 | 28.95/0.8994 | 3.0118 | 32.13/0.9353 | 6.0000 | 29.97/0.8821 | 1.8591 |
| flower | 34.51/0.9421 | 1.0908 | 36.10/0.9465 | 1.0908 | 34.46/0.9300 | 3.0000 | 32.56/0.9297 | 1.5815 | 28.68/0.6884 | 1.2466 | 35.86/0.9613 | 2.5087 | 35.13/0.9487 | 6.0000 | 37.00/0.9397 | 1.3162 |
| flowers | 30.31/0.9178 | 1.3628 | 32.15/0.9284 | 1.3628 | 31.25/0.9107 | 3.0000 | 30.39/0.9145 | 2.2879 | 25.56/0.7406 | 1.2368 | 31.47/0.9400 | 2.7691 | 32.31/0.9351 | 6.0000 | 33.31/0.9195 | 1.9729 |
| girl | 33.68/0.9040 | 1.2112 | 34.83/0.9115 | 1.2112 | 33.60/0.8977 | 3.0000 | 31.00/0.8934 | 1.9572 | 28.05/0.7143 | 1.2471 | 34.26/0.9359 | 2.5688 | 36.20/0.9475 | 6.0000 | 34.88/0.9086 | 1.5742 |
| goldhill | 32.10/0.8662 | 1.1928 | 32.90/0.8721 | 1.1928 | 31.85/0.8613 | 3.0000 | 30.08/0.8839 | 2.2479 | 27.39/0.7058 | 1.2467 | 33.04/0.9124 | 2.7309 | 33.36/0.9251 | 6.0000 | 32.76/0.8738 | 1.6784 |
| lenna | 32.74/0.8698 | 1.2330 | 33.91/0.8847 | 1.2330 | 32.45/0.8537 | 3.0000 | 30.30/0.8667 | 1.8882 | 27.11/0.6725 | 1.2448 | 32.86/0.9028 | 2.6163 | 33.93/0.9271 | 6.0000 | 33.05/0.8630 | 1.3779 |
| man | 29.48/0.8517 | 1.5129 | 30.79/0.8606 | 1.5129 | 29.38/0.8467 | 3.0000 | 28.55/0.8714 | 2.8191 | 23.00/0.6529 | 1.2359 | 27.80/0.8675 | 3.0622 | 28.43/0.8790 | 6.0000 | 28.00/0.8150 | 2.2676 |
| pens | 33.19/0.9185 | 1.2515 | 34.28/0.9223 | 1.2515 | 33.22/0.9133 | 3.0000 | 30.57/0.9040 | 2.0352 | 27.69/0.7356 | 1.2443 | 33.840.9455 | 2.5939 | 33.36/0.9336 | 6.0000 | 35.48/0.9296 | 1.6637 |
| pepper | 33.02/0.8672 | 1.2327 | 34.35/0.8825 | 1.2327 | 32.68/0.8513 | 3.0000 | 30.16/0.8651 | 1.8408 | 26.26/0.6611 | 1.2450 | 31.90/0.8905 | 2.5419 | 33.39/0.9164 | 6.0000 | 33.06/0.8493 | 1.3594 |
| sailboat | 30.84/0.8580 | 1.3635 | 31.83/0.8687 | 1.3635 | 30.69/0.8509 | 3.0000 | 29.27/0.8872 | 2.5109 | 24.69/0.6825 | 1.2369 | 29.74/0.8827 | 2.9071 | 30.79/0.9049 | 6.0000 | 30.71/0.8437 | 1.9317 |
| tiffany | 30.23/0.8797 | 1.5177 | 31.45/0.8869 | 1.5177 | 26.35/0.8325 | 3.0000 | 30.95/0.8941 | 1.6797 | 26.79/0.6619 | 1.2458 | 32.45/0.9121 | 2.5277 | 35.53/0.9388 | 6.0000 | 33.28/0.8641 | 1.2512 |
| yacht | 31.92/0.9184 | 1.2624 | 32.88/0.9176 | 1.2624 | 31.88/0.9139 | 3.0000 | 31.07/0.9218 | 2.0722 | 25.51/0.7157 | 1.2375 | 31.68/0.9468 | 2.7963 | 34.41/0.9575 | 6.0000 | 34.76/0.9346 | 1.6890 |
| Lichtenstein | 32.81/0.9170 | 0.9616 | 33.40/0.9189 | 0.9616 | 32.65/0.9096 | 3.0000 | 31.78/0.9416 | 1.9060 | 26.47/0.6537 | 1.2431 | 32.11/0.9355 | 2.7059 | 33.24/0.9454 | 6.0000 | 32.05/0.9009 | 1.4112 |
| airplane | 33.27/0.9249 | 1.0420 | 34.31/0.9281 | 1.0420 | 33.09/0.9132 | 3.0000 | 31.88/0.9322 | 1.7799 | 25.48/0.6683 | 1.2398 | 31.25/0.9414 | 2.6279 | 32.80/0.9356 | 6.0000 | 33.50/0.9101 | 1.5224 |
| cameraman | 34.06/0.9403 | 0.9790 | 35.01/0.9376 | 0.9790 | 33.81/0.9328 | 3.0000 | 32.28/0.9393 | 1.6638 | 26.83/0.6960 | 1.2417 | 33.30/0.9606 | 2.5752 | 36.33/0.9695 | 6.0000 | 35.72/0.9372 | 1.3104 |
| kodim05 | 29.31/0.9009 | 1.7361 | 29.99/0.9057 | 1.7361 | 28.90/0.8981 | 3.0000 | 28.07/0.9264 | 3.1169 | 20.93/0.6831 | 1.2170 | 26.06/0.8883 | 3.2515 | 26.44/0.8695 | 6.0000 | 27.49/0.8659 | 2.9456 |
| kodim09 | 33.96/0.9138 | 1.0587 | 34.94/0.9209 | 1.0587 | 33.51/0.8938 | 3.0000 | 32.12/0.9198 | 1.5868 | 25.52/6301 | 1.2404 | 31.66/0.9308 | 2.5635 | 33.01/0.9373 | 6.0000 | 33.31/0.8975 | 1.2127 |
| kodim14 | 30.35/0.8569 | 1.4402 | 30.72/0.8604 | 1.4402 | 30.21/0.8550 | 3.0000 | 28.76/0.8959 | 2.8352 | 23.72/0.6868 | 1.2375 | 28.74/0.8844 | 3.0403 | 29.63/0.8913 | 6.0000 | 29.26/0.8396 | 2.3378 |
| kodim15 | 32.56/0.8555 | 1.2281 | 33.03/0.8569 | 1.2281 | 30.54/0.8463 | 3.0000 | 30.21/0.8736 | 2.0388 | 26.51/0.6699 | 1.2417 | 31.93/0.8980 | 2.6657 | 33.01/0.9215 | 6.0000 | 31.80/0.8412 | 1.4603 |
| kodim20 | 33.71/0.9269 | 0.9231 | 34.39/0.9292 | 0.9231 | 22.30/0.9195 | 3.0000 | 31.64/0.9449 | 1.7675 | 24.67/0.6616 | 1.2418 | 30.54/0.9374 | 2.6004 | 31.42/0.9304 | 6.0000 | 32.18/0.9084 | 1.3365 |
| kodim21 | 31.13/0.8988 | 1.1979 | 31.74/0.9001 | 1.1979 | 30.10/0.8855 | 3.0000 | 30.32/0.9239 | 2.2154 | 24.01/0.6409 | 1.2375 | 29.20/0.9105 | 2.8870 | 30.07/0.9209 | 6.0000 | 29.89/0.8836 | 1.8190 |
| kodim23 | 34.84/0.9216 | 1.1083 | 35.51/0.9242 | 1.1083 | 33.04/0.9022 | 3.0000 | 31.96/0.9137 | 1.4874 | 26.31/0.6489 | 1.2416 | 32.54/0.9413 | 2.4708 | 35.15/0.9559 | 6.0000 | 34.39/0.9090 | 1.1981 |
| milkdrop | 36.22/0.9044 | 0.9991 | 36.86/0.9041 | 0.9991 | 35.11/0.8870 | 3.0000 | 32.21/0.8851 | 1.3529 | 27.79/0.6495 | 1.2467 | 34.38/0.9316 | 2.3724 | 35.93/0.9351 | 6.0000 | 35.79/0.8819 | 1.0110 |
| **average** | **32.21/0.8969** | **1.2467** | **33.16/0.9020** | **1.2467** | **31.28/0.8865** | **3.0000** | **30.66/0.9079** | **2.1040** | **25.64/0.6780** | **1.2402** | **31.34/0.9185** | **2.7439** | **32.66/0.9273** | **6.0000** | **32.51/0.8861** | **1.6769** |

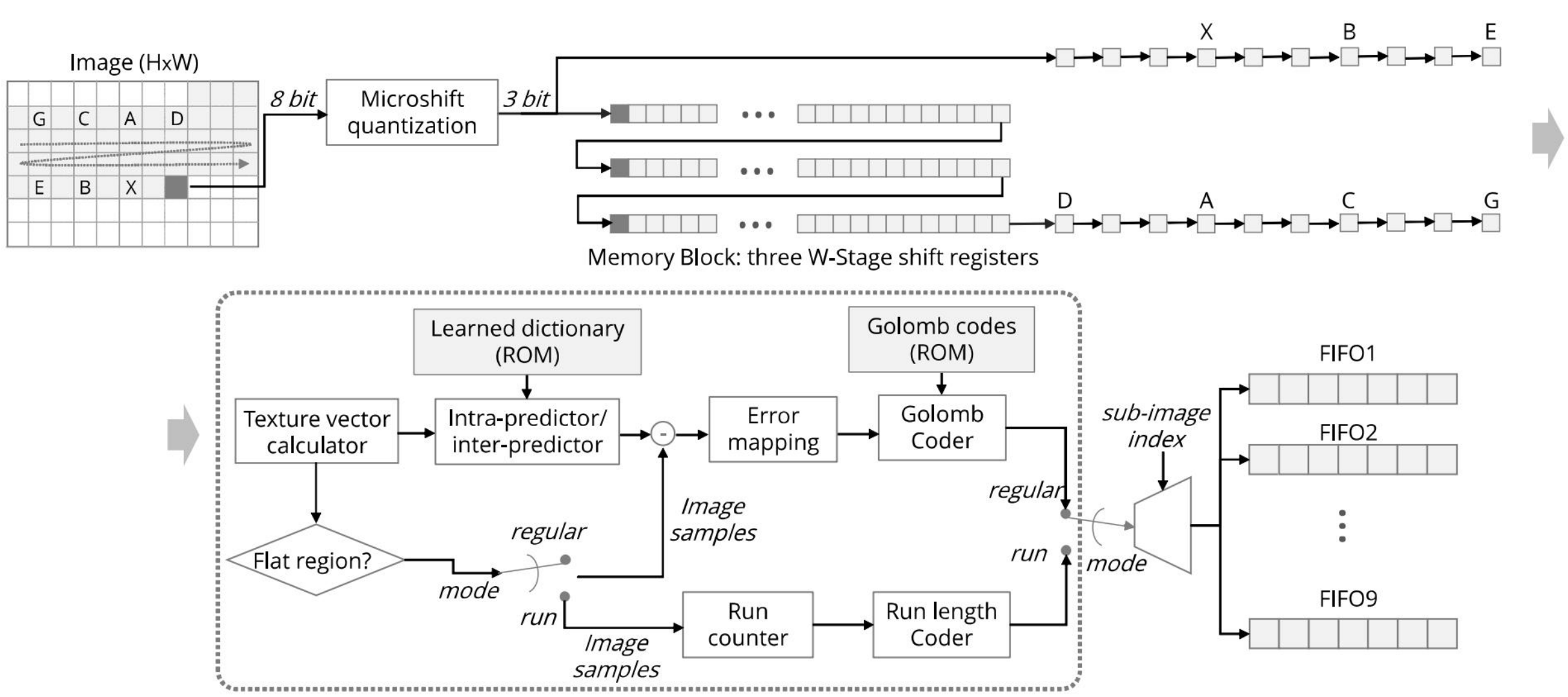

Fig. 10. The diagram of hardware implementation.

TABLE IV
HARDWARE RESOURCES

| Function | The *Microshift* compression |
|---|---|
| Adders/subtractors | 171 |
| Shift operators | 47 |
| Multiplexers | 459 |
| RAM | 9 kbit (640×480);18 kbit (1280×720) |

Having obtained the template pixels, the texture vector can be calculated as Eq. 10 and be used to distinguish whether the local patch is uniform or not. For local patches, the compression will go into the runlength mode; otherwise the pixel will be compressed through intra- or inter-prediction according to the subimage index. The learned predictors for all the 313 contexts (Eq. 12) are stored in the read-only memory (Fig. 10). Then the prediction residues will be mapped through Eq. 14-16 and then encoded using the Golomb codes pre-stored in the memory. The memory in the system can be efficiently implemented through FPGA embedded block memory. Furthermore, because the transmission from the image sensor to the compression circuit is serial and each in clock cycle one pixel is processed, we built the system through pipeline for efficient computation and better scalability. The overhead latency due to the pipeline is 8 clocks and it takes $(H \times W + 8)$ clock cycles to compress the entire image.

After compression of each pixel, the bitstream with variable length will serve as input to the FIFOs. Because our method is designed for progressive compression, we need nine FIFOs to buffer the compressed bitstream from the corresponding subimage so that the data for different subimages can be stored and transmitted serially.

Finally, it should be noted that all the operators in the implementation are hardware friendly. Table IV summarizes the resource utilization of the compression circuit. Only simple adders/subtractors or shifters are needed. Furthermore, our method is efficient in memory usage, because during the raster scan there is no need to store the whole image for compression. As shown in Table IV, the hardware is scalable to different image resolutions: the logic utilization remains the same and the memory utilization is linearly proportional to the image width. In order to transmit the subimages progressively, we also need the memory (output FIFOs in Fig. 10) to store the compressed bitstream. When the progressive compression is not required for power saving, even this storage can be saved.

### B. FPGA implementation

We implement the *Microshift* on a Terasic DE1 FPGA board which uses Altera Cyclone V chip[1]. A Terasic D8M camera[2] is used for image acquisition. The resolution of the image sensor is configured to 640×480 ($W = 640$ in Fig. 10). The image captured by the camera is fed to a monitor for display through the VGA port of the FPGA board and serves as a reference image without compression. On the other hand, each pixel of the image will be serially compressed and the compressed bitstream is then transmitted to a PC through UART protocol. The image is progressively reconstructed on a PC and the decompression result is shown on another monitor. The photo of this demonstration system is shown in Fig. 11.

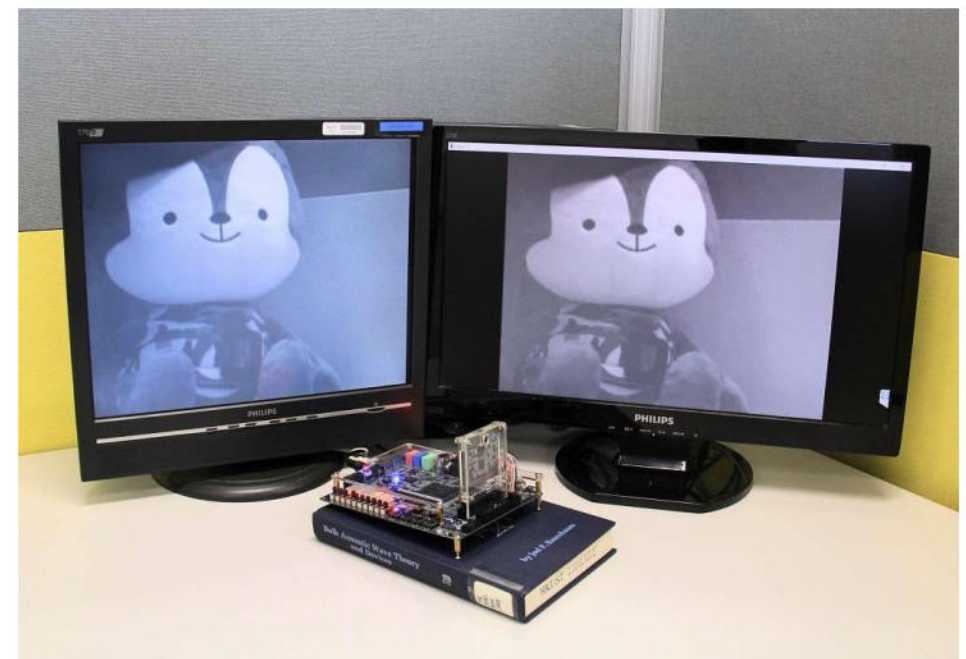

Fig. 11. FPGA demonstration system for the proposed *Microshift*.

TABLE V
FPGA RESOURCE UTILIZATION FOR THE COMPRESSION CORE

| FPGA board | Altera Cyclone V (5CSEMA5F31C8) |
|---|---|
| Logic utilization (in ALMs) | 1,154 / 32,070 (4%) |
| Combinational ALUTs | 1,947 |
| Dedicated logic registers | 1006 / 64,140 (<1%) |
| Block memory bits | 9,216 / 4,065,280 (<1%) |
| Operating frequency | 50 MHz |
| Estimated dynamic power | 1.34 mW |

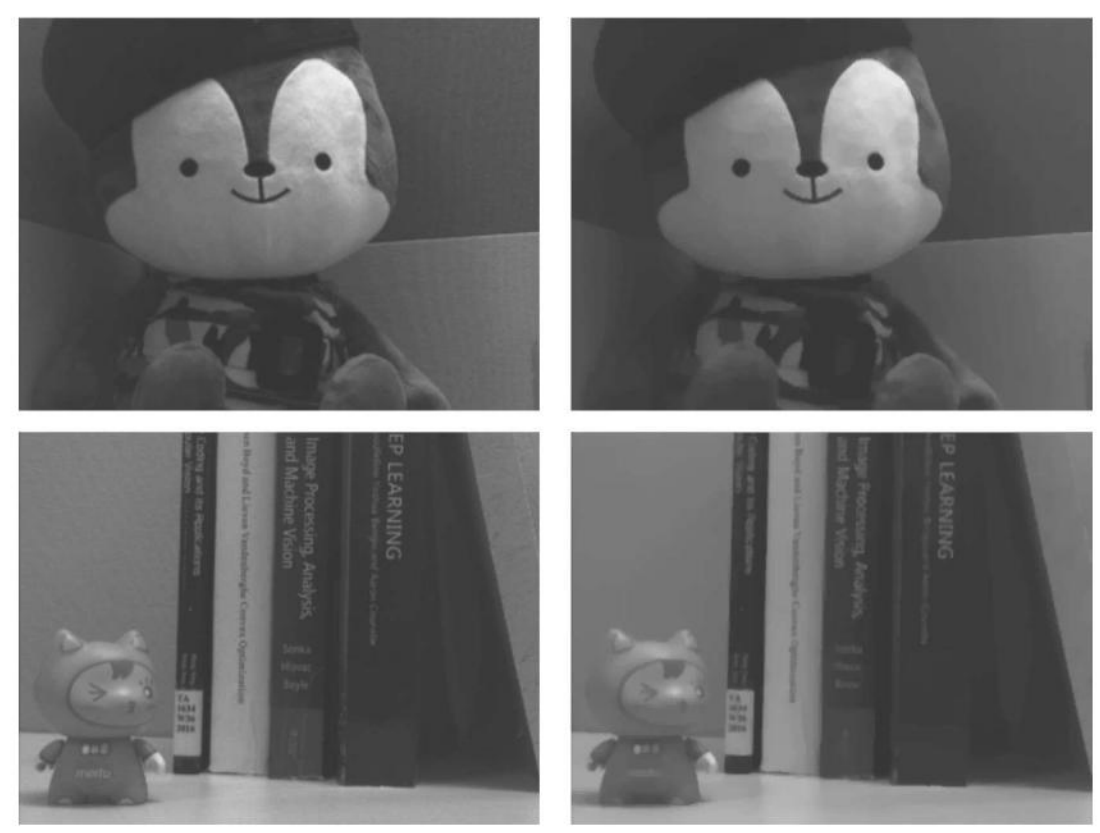

Fig. 12. Real images captured from the FPGA demo system. Left: raw images. Right: corresponding decompressed images.

Table V summarizes the FPGA resource utilization. Because of the algorithm hardware co-design methodology, the implementation is efficient: the logic utilization is 4%, and the block memory (M10k) utilization is less than 1%. Figure 12 shows the results captured by our demo system. Both the raw image captured by the image sensor and the compressed image using our method are shown. Here, we use the MRF model to maximize the decompressed image quality. It can be seen that the edges in the decompression image are sharp and objects can be clearly distinguished. The overall decompression result exhibits good visual quality, which is suitable for sensing applications.

[1]Terasic DE1 board: http://de1.terasic.com.tw/
[2]Terasic D8M camera: http://d8m.terasic.com.tw/

TABLE VI
CHARACTERISTIC OF OUR VLSI DESIGN

| | | | |
|---|---|---|---|
| Technology | Global Foundry 0.18um process | | |
| Function | Microshift image compression | | |
| Operation frequency | 100 MHz | | |
| Resolution | 256×256 | 640×480 | 1280×720 |
| Cell area | 0.45 $mm^2$ | 0.82 $mm^2$ | 1.48 $mm^2$ |
| Equivalent gate count | 45.5 K | 82.9 K | 129.6 K |
| Memory usage | 6.5 kbit | 9 kbit | 18 kbit |

a

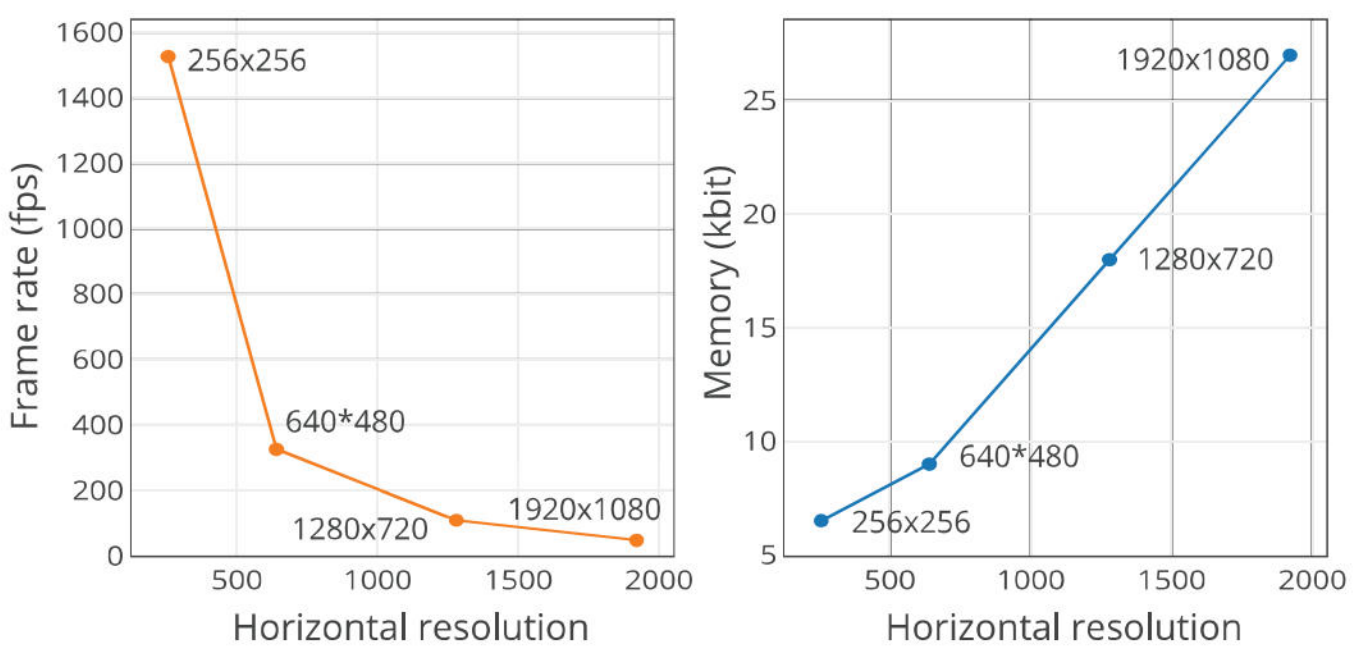

Fig. 13. Performance and memory utilization for different image resolutions.

## C. ASIC implementation

To demonstrate the energy efficiency of our design, we also synthesize our algorithm to ASIC implementation using a GlobalFoundry 0.18 $\mu$m process. Table VI summarizes the major characteristics of our VLSI design, and all the results are based on gate-level synthesis and simulation (Synopsys Design Compiler, Mentor Modelsim, Synopsys Power Compiler and Cadence Encounter). The power is optimized through clock gating and operand isolation. The total cell area for different image resolutions is also reported in the table[1].

Table VII compares the design with other on-chip compression implementations in the literature. Since our method is fully compatible with the mainstream APS image sensor, and it does not affect the pixel design and the fill factor, the image sensor is capable of high quality image acquisition. Also, due to the raster scan compression manner, the design provides a high throughput. On one hand, our method achieves high compression performance, and provides high image quality compared to other on-chip compression methods. On the other hand, our circuit is power efficient. The table gives the measured power for the work of VPIC, compressive sensing, DCT and lossless prediction, and here we give the estimated power for our work. For fair comparison, we use the power figure of merit (FOM) which is defined as power consumption normalized to the frame rate and the number of pixels. The power FOM for this work is 19.7 pW/pixel·frame at the working voltage 1V. The power FOM for the pixel array is estimated by 40 pW/pixel·frame (typical figure according to [31], [45], so the total power FOM is 59.7 pW/pixel·frame by estimation. Furthermore, our method shows the advantage of good scalability as shown in Fig. 13. Finally, the feature of progressive compression makes our method even more appealing to low-power wireless sensing applications.

[1]Here all the resource utilization number just includes the compression block, and excludes the output FIFOs that store the compressed bitstream.

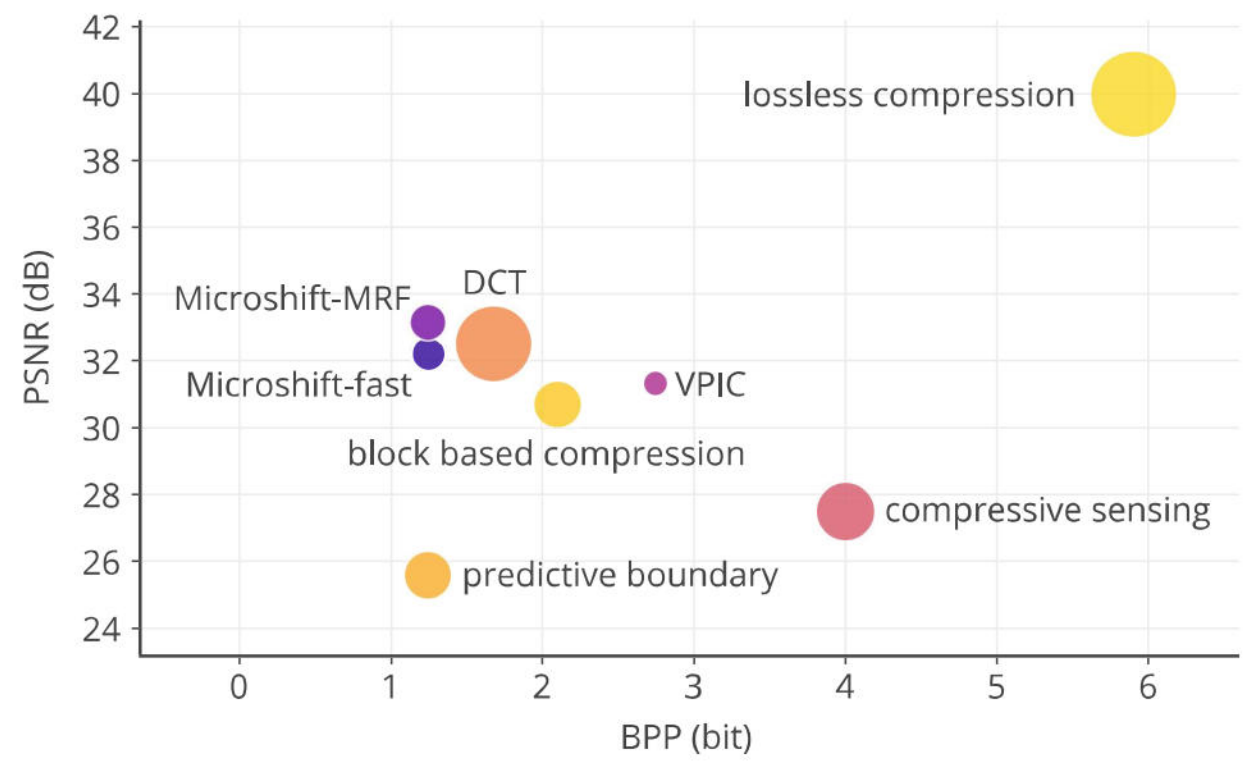

Fig. 14. Comparison of various on-chip compression methods in terms of PSNR (dB), bit per pixel (BPP) and power FOM (pW/pixel·frame). The area for each scatter point denotes the power FOM in logarithm scale. Compared with other methods, our work demonstrates significant advantage in terms of these three measures.

Fig. 14 gives an intuitive comparison in terms of BPP, PSNR and power FOM. In this figure, the area for each scatter point denotes the power FOM in logarithm scale. Since the predictive boundary method [28] does not report the frame rate and the blocked based compression [27] does not estimate the total power consumption, the point area in Fig. 14 for these two works does not reflect the power FOM accurately. However, their power consumption is assumed to be much larger than this work, because both of them build on quadtree decomposition which is more computationally intensive than raster scan. Though the estimated power FOM is larger than VPIC [31], our work is an off-array processor and fully compatible to typical image sensors. Besides, this work provides higher compression ratio and can significantly reduce the transmission power, which accounts for most of the power in the WSNs [8]. Furthermore, our work exhibits good decompression quality (PSNR>33dB and BPP=1.1) while the VPIC is not suitable for high quality imaging (PSNR=20 when BPP=1 in [31]). In all, Fig. 14 shows that our method achieves the best compression performance while maintaining relatively low power, outperforming other methods by a large margin.

## VII. CONCLUSION

In this paper, we propose the *Microshift* based on an algorithm-hardware co-design methodology, which achieves good compression performance while preserving hardware friendliness. Then we propose two decompression methods, the more efficient FAST method, and the higher quality MRF method. Both methods can reconstruct images progressively. The compression performance is validated through extensive experiments. Finally, we propose a hardware implementation architecture, and demonstrate the prototype system using FPGA. We compare our efficient VLSI implementation with previous work, validating that our method is competitive for low power wireless sensing applications. Our algorithm and hardware implementation are freely available for public reference.

TABLE VII
COMPARISON WITH ON-CHIP COMPRESSION WORKS

| Method | Compression scheme | **This work** | Block based [27] | Predictive boundary [28] | VPIC [31] | Compressive sensing [34] | DCT [20] | Lossless predictive [15] |
|---|---|---|---|---|---|---|---|---|
| Process | Technology (um) | **0.18** | 0.18 | 0.35 | 0.18 | 0.15 | 0.5 | 0.35 |
| Image sensor | Architecture | **off-array processor** | pixel level & off-array | pixel level & off-array | column level | column level | pixel level | off-array processor |
| | Resolution | **256×256** | 64×64 | 64×64 | 816×640 | 256×256 | 104×128 | 80×44 |
| | Pixel strcuture | **APS** | DPS | DPS | APS | APS | APS | APS |
| | Pixel pitch (um^2) | **5.6** | 14 | 39 | 1.85 | 5.5 | 13.5 | 30 |
| | Fill factor | **>40%** | 15% | 12% | 13 | N/A | 46% | 18% |
| | Area (mm^2) | **1.2×1.2 (processor)** | 0.985×0.952 | 2.2×0.25 (processor) | 2.16×1.36 | 2.9×3.5 | 2.4×1.8 | 2.6×6.0 |
| Resource | Memory content | **bitstream** | quadtree | quadtree | none | none | none | none |
| Performance | Frame rate (fps) | **1530 (max.)** | N/A | N/A | 111 | 1920 | 25 | 435 |
| | Throughput (Mp/s) | **100 (max.)** | N/A | N/A | 58 | 7.9 | 0.3 | 1.5 |
| | Power FOM (pW/pixel•frame) | **19.7 (processor)** | N/A | N/A | 13.9 | 765 | 6010 | 21973 |
| | Bit per pixel (BPP) | **1.1** | 2.1 | 1.2 | 2.7 | 6.2 | 1.6 | 5.9 |
| | PSNR (dB) | **33.2** | 30.7 | 25.6 | 31.3 | 33.2 | 32.5 | >60 |
| Feature | Scalability | **$O$(W)** | $O$(HW) | $O$(HW) | $O$(HW) | $O$(HW) | $O$(HW) | $O$(W) |
| | Progressive reconstruction | **yes** | no | no | no | yes | yes | no |

Further improvements of our compression scheme is promising. First, deep neural networks can be employed to learn the mapping from the microshift pattern and the ground truth, so the decompression is simply the per-pixel prediction through a forward network computation. Second, the microshift sub-quantization can be implemented in the analog domain [46]. In this way, since data redundancy is compressed in the sensory front end more power savings can be expected, which is appealing to WVSNs.

## ACKNOWLEDGMENT

The authors would like to acknowledge the financial support from HK Innovation and Technology Fund ITF, Grant ITS/211/16FP.

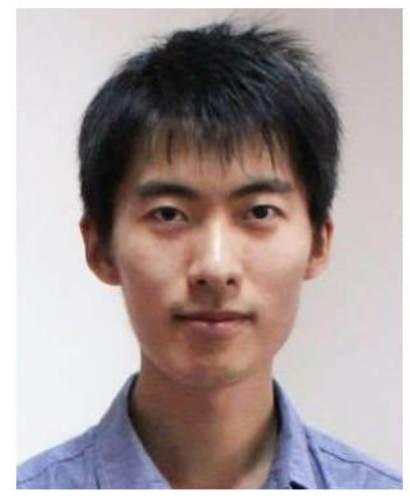

**Bo Zhang** received the B.Eng. degree of optical engineering from Zhejiang University, Zhejiang, China in 2013. He is currently pursuing the Ph.D. degree with the Department of Electronic and Computer Engineering at the Hong Kong University of Science and Technology. His research interests include image processing and computational photography.

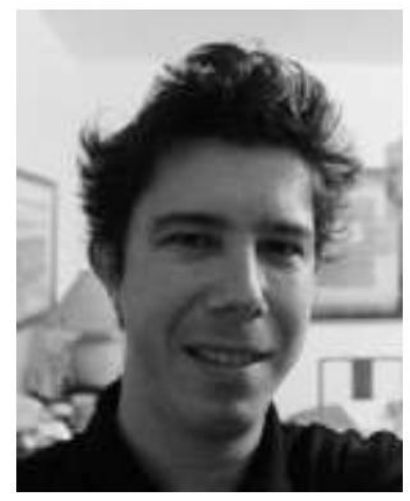

**Pedro V. Sander** is an Associate Professor in the Department of Computer Science and Engineering at the Hong Kong University of Science and Technology. His research interests lie mostly in real-time rendering, graphics hardware, and geometry processing. He received a Bachelor of Science in Computer Science from Stony Brook University in 1998, and Master of Science and Doctor of Philosophy degrees from Harvard University in 1999 and 2003, respectively. After concluding his studies, he was a member of the Application Research Group of ATI Research, where he conducted real-time rendering and general-purpose computation research with latest generation and upcoming graphics hardware. In 2006, he moved to Hong Kong to join the Faculty of Computer Science and Engineering at The Hong Kong University of Science and Technology.

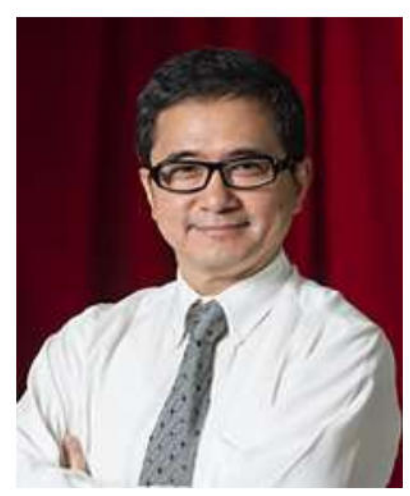

**Chi-Ying Tsui** received the B.S. degree in electrical engineering from the University of Hong Kong, Hong Kong, and the Ph.D. degree in computer engineering from the University of Southern California, Los Angeles, CA, USA, in 1994. He joined the Department of Electronic and Computer Engineering, Hong Kong University of Science and Technology, Hong Kong, in 1994, where he is currently a Full Professor.

He has published more than 160 referred publications and holds ten U.S. patents on power management, VLSI, and multimedia systems. His current research interests include designing VLSI architectures for low power multimedia and wireless applications, developing power management circuits and techniques for embedded portable devices, and ultra-low power systems.

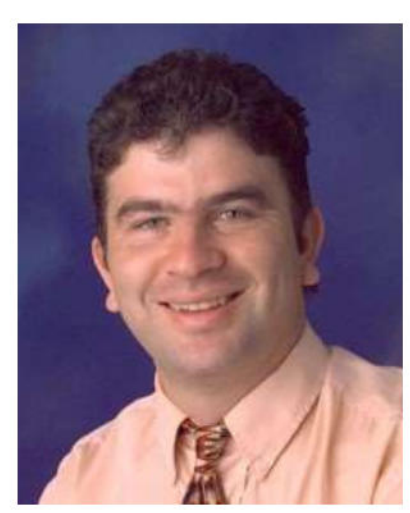

**Amine Bermak** received the Masters and PhD degrees, both in electrical and electronic engineering (Microelectronics and Microsystems), from Paul Sabatier University, Toulouse, France in 1994 and 1998, respectively. In November 1998, he joined Edith Cowan University, Perth, Australia as a research fellow working on smart vision sensors. In January 2000, he was appointed Lecturer and promoted to Senior Lecturer in December 2001. In July 2002, he joined the Electronic and Computer Engineering Department of Hong Kong University of Science and Technology (HKUST), where he held a full Professor position and ECE Associate Head for Research and Postgraduate studies. He has also been serving as the Director of Computer Engineering as well as the Director of the Master Program in IC Design. He is also the founder and the leader of the Smart Sensory Integrated Systems Research Lab at HKUST. Currently, Prof. Bermak is with Hamad Bin Khalifa University, Qatar Foundation, Qatar and has been holding a full professor appointment as well as acting Associate Provost.